\documentclass[11pt,draftcls,onecolumn]{IEEEtran}

\usepackage{color}
\usepackage{amscd}
\usepackage{amsmath}
\usepackage{enumerate}
\usepackage{theorem}
\usepackage{cite}
\usepackage{stfloats}
\usepackage{epsfig}
\usepackage{verbatim}
\usepackage{epsfig}
\usepackage{graphicx}
\usepackage{amssymb}

\theoremstyle{plain} \theorembodyfont{\itshape}
\theoremheaderfont{\scshape}

\theorembodyfont{\itshape} \theoremheaderfont{\scshape}

\theoremstyle{plain} \theorembodyfont{\itshape}
\theoremheaderfont{\scshape}

\begin{document}

\title{Blind Pilot Decontamination}

\author{
Ralf~R.~M{\"u}ller$^\ast$,~\IEEEmembership{Senior Member,~IEEE}  and Laura Cottatellucci and Mikko Vehkaper\"a%
\thanks{Manuscript submitted Sep 25, 2013.
This paper was presented in part at the 2013 ITG Workshop on Smart Antennas in Stuttgart, Germany, the 2013 IEEE
Vehicular Technology Conference in Dresden, Germany, and the 47th Asilomar Conference on Signals, Systems, and Computers in Pacific Grove, CA, USA.}
\thanks{R. R.\ M{\"u}ller is with Friedrich-Alexander Universit\"at Erlangen-N\"urnberg, Cauerstr.~7/LIT, 91058 Erlangen, Germany, and The Norwegian University of Science and Technology, Trondheim, Norway,
(e-mail: mueller@lnt.de). L.\ Cottatellucci is with Institute Eurecom, Sophia-Antipolis, France, (e-mail:cottatellucci@eurecom.fr).
M.\ Vehkaper\"a is with Aalto University, Helsinki, Finland, and Royal Institute of Technology, Stockholm, Sweden (e-mail: mikkov@kth.se).
}%
\thanks{This work was supported in part by the European Commission via the 7th framework project HARP.}
}

\markboth{Submitted to IEEE Journal of Selected Topics in Signal Processing}{M\"{u}ller et al.: Blind Pilot Decontamination}

\maketitle
\begin{abstract}
A subspace projection to improve channel estimation in massive multi-antenna systems is proposed and analyzed. 
Together with power-controlled hand-off, it can mitigate the pilot contamination problem without the need for coordination among cells.
The proposed method is blind in the sense that it does not require pilot data to find the appropriate subspace.
It is based on the theory of large random matrices that predicts that the eigenvalue spectra of large sample covariance matrices can asymptotically decompose into disjoint bulks as the matrix size grows large.
Random matrix and free probability theory are utilized to predict under which system parameters such a bulk decomposition takes place. Simulation results are provided to confirm that the proposed method outperforms conventional linear channel estimation if bulk separation occurs.

\end{abstract}


\newcommand{\fzv}[2]{\noindent #1 \hfill \parbox{13.2cm}{#2}}
\def\mathlette#1#2{{\mathchoice{\mbox{#1$\displaystyle #2$}}%
                              {\mbox{#1$\textstyle #2$}}%
                              {\mbox{#1$\scriptstyle #2$}}%
                              {\mbox{#1$\scriptscriptstyle #2$}}}}
\newcommand{\matr}[1]{\mathlette{\boldmath}{#1}}
\newcommand{\SINR}{{sig\-nal--to--dis\-tor\-tion ratio}}
\newcommand{\SNR}{{sig\-nal--to--noise ratio}}
\renewcommand{\j}{{\rm j}}
\newcommand{\RR}{\mathbb{R}}
\newcommand{\CC}{\mathbb{C}}
\newcommand{\NN}{\mathbb{N}}
\newcommand{\ZZ}{\mathbb{Z}}
\newcommand{\pic}{\pi}
\newcommand{\deltaf}{\delta}
\def\argmin{\mathop{\rm argmin}}
\def\argmax{\mathop{\rm argmax}}
\newcommand{\diag}{{\rm diag}}
\def\expect{\mathop{\mbox{$\mathsf{E}$}}}
\newcommand{\hermite}[1]{{#1^{\rm H}}}
\newcommand{\transp}[1]{{#1^{\rm T}}}
\newcommand{\conj}[1]{{#1^{\ast}}}
\newcommand{\e}{{\rm e}}
\newcommand{\iu}{{\rm j}}
\newcommand{\vnull}{{\rm \bf 0}}
\newcommand{\I}{{\rm \bf I}}
\newcommand{\energy}{{E}}
\newcommand{\prob}[2]{{\rm p}_{#1}\!\!\left( #2 \right) }
\newcommand{\Prob}[2]{{\rm P}_{#1}\!\!\left( #2 \right) }
\newcommand{\proba}[2]{\breve{\rm p}_{#1}\!\!\left( #2 \right) }
\newcommand{\Proba}[2]{\breve{\rm P}_{#1}\!\!\left( #2 \right) }
\newcommand{\Probi}[2]{{\rm P}_{#1}^{-1}\!\left( #2 \right) }
\newcommand{\RT}[2]{{\rm R}_{#1}\!\left( #2 \right) }
\newcommand{\Q}{{\rm Q}}
\newcommand{\load}{{\beta}}
\newcommand{\sign}{{\rm sign}}
\newcommand{\dirac}[1]{\deltaf \! \left( #1 \right)}
\newcommand{\kron}[1]{\deltaf \! \left[ #1 \right]}
\newcommand{\MomGen}[2]{{\Phif}_{#1}\left( #2 \right) }
\newcommand{\define}{\stackrel{\triangle}{=}}
\newcommand{\D}{\displaystyle}
\newcommand{\eq}[1]{(\ref{#1})}
\newcommand{\eqs}[2]{(\ref{#1}) and (\ref{#2})}
\newcommand{\eqd}[3]{(\ref{#1}), (\ref{#2}), and (\ref{#3})}
\newcommand{\eqv}[4]{(\ref{#1}), (\ref{#2}), (\ref{#3}), and (\ref{#4})}
\newcommand{\alphabet}{{\cal B}}
\newcommand\ie{{\textsl{i.e.\,}}}
\newcommand\eg{{\textsl{e.g.\,}}}

\newcommand\va{{\bf a}} 
\newcommand\vb{{\bf b}}
\newcommand\vc{{\bf c}}
\newcommand\vd{{\bf d}}
\newcommand\ve{{\bf e}}
\newcommand\vf{{\bf f}}
\newcommand\vg{{\bf g}}
\newcommand\vh{{\bf h}}
\newcommand\vi{{\bf i}}
\newcommand\vj{{\bf j}}
\newcommand\vk{{\bf k}}
\newcommand\vl{{\bf l}}
\newcommand\vm{{\bf m}}
\newcommand\vn{{\bf n}}
\newcommand\vo{{\bf o}}
\newcommand\vp{{\bf p}}
\newcommand\vq{{\bf q}}
\newcommand\vr{{\bf r}}
\newcommand\vs{{\bf s}}
\newcommand\vt{{\bf t}}
\newcommand\vu{{\bf u}}
\newcommand\vv{{\bf v}}
\newcommand\vw{{\bf w}}
\newcommand\vx{{\bf x}}
\newcommand\vy{{\bf y}}
\newcommand\vz{{\bf z}}
\newcommand\mA{{\bf A}} 
\newcommand\mB{{\bf B}}
\newcommand\mC{{\bf C}}
\newcommand\mD{{\bf D}}
\newcommand\mE{{\bf E}}
\newcommand\mF{{\bf F}}
\newcommand\mG{{\bf G}}
\newcommand\mH{{\bf H}}
\newcommand\mI{{\bf I}}
\newcommand\mJ{{\bf J}}
\newcommand\mK{{\bf K}}
\newcommand\mL{{\bf L}}
\newcommand\mM{{\bf M}}
\newcommand\mN{{\bf N}}
\newcommand\mO{{\bf O}}
\newcommand\mP{{\bf P}}
\newcommand\mQ{{\bf Q}}
\newcommand\mR{{\bf R}}
\newcommand\mS{{\bf S}}
\newcommand\mT{{\bf T}}
\newcommand\mU{{\bf U}}
\newcommand\mV{{\bf V}}
\newcommand\mW{{\bf W}}
\newcommand\mX{{\bf X}}
\newcommand\mY{{\bf Y}}
\newcommand\mZ{{\bf Z}}

\section{Introduction}\label{sec_intro}
\IEEEPARstart{I}{n} \cite{marzetta:06}, a multiple antenna system was proposed that mimics the idea of spread-spectrum.
Like a large processing gain can be realized in a spread-spectrum system by massive use of radio spectrum, a large array gain is realized by a massive use of antennas elements. This system design has attracted considerable attention recently, see e.g.\ \cite{rusek:12} for a survey. It is commonly referred to as {\em massive MIMO}. Its advantage over the old spread-spectrum idea lies in the fact that antennas can be manufactured in arbitrarily high numbers, while radio spectrum is limited.

Given perfect channel state information, the signals received at all antenna elements can be combined coherently.
The array gain grows unboundedly with the number of antenna elements at the access point. Therefore, massive use of antennas elements can overcome both multiuser interference and thermal noise for any given number of users and any given powers of the interfering users.

Given a rich scattering environment, the access point is able to concentrate the radiated energy within an area around the designated receiver that is only a fraction of a squared wavelength \cite{rusek:12}. 
This is in contrast to earlier methods of array processing in wireless communications based upon angles of arrival and departure which are limited to concentrate their energy into a certain direction rather than an area limited in both angular and radial domain.

Since power in the far-field can be amplified (in theory) beyond limits by means of array gains, the ultimate limit of a single-cell massive MIMO system is only constrained by the coherence time of the channel \cite{marzetta:10}. The number of terminals needs to stay below the coherence time to have sufficient degrees of freedom for channel estimation. In practice, the coherence time needs to be significantly larger than the number of terminals to also allow for data transmission in addition to pilot symbols for channel estimation.

In \cite{marzetta:10}, however, a pessimistic conclusion about the performance of massive MIMO in {\em multi-cellular} systems was reached.
Based on the explicit assumption of no coordination among cells and on the implicit assumption of linear channel estimation \cite[Eq.\ (5)]{marzetta:10}, it was concluded that the array gain can be achieved only for data detection, but not for channel estimation. The author argued that channel state information, though not required to be perfect, must have at least a certain quality in order to utilize unlimited array gains. 
As a result, pilot interference from neighboring cells would limit the ability to obtain sufficiently accurate channel estimates
and be the new bottleneck of the system.
This effect, commonly referred to as {\em pilot contamination} \cite{jose:11}, is treated as a fundamental effect in many works, e.g.\ \cite{marzetta:10,jose:11,ngo:13,gopalakrishnan:11,fernandes:12,krishnan:12}.

Recent works have indicated that pilot contamination may not be as fundamental as it was thought to be:
Using Bayesian channel estimation, \cite{yin:12} found that pilot contamination can vanish under certain conditions on the channel covariance matrix if some cooperation among cells is allowed for.
Using an eigenvalue decomposition of the sample covariance matrix of the received signal, \cite{ngo:12} found that for a wide range of system parameters, the channel can be estimated with greater accuracy than with linear methods.

In early conference versions of this work \cite{mueller:13a,cottatellucci:13}, we showed that pilot contamination is not a fundamental limitation for massive MIMO systems, if the coherence time is not smaller than the number of antennas at the base station and a power margin between users of interest and interfering users can be provided, e.g.\ by means of path loss. 
In that case the array gain can be utilized to have the accuracy of channel estimation growing unboundedly with the number of antennas at polynomial complexity.
As in \cite{ngo:12}, the approach in \cite{mueller:13a,cottatellucci:13} starts with an eigenvalue decomposition of the sample covariance matrix (or equivalently a singular value decomposition of the received signal matrix). Unlike \cite{ngo:12}, it does not aim to subsequently estimate the channel matrix before performing data detection. It projects the received signal onto an (almost) interference-free subspace where communication is governed by a non-linear compound channel that can be estimated more easily.

The work in \cite{mueller:13a,cottatellucci:13} has received criticism for the assumption that the coherence time is assumed to be larger than the number of base station antennas. In this journal version of our work, we remove this assumption and also allow for coherence times shorter than the number of base station antennas.
We find that the main conclusions of \cite{mueller:13a,cottatellucci:13} remain valid and show that our proposed method of channel estimation based on subspace projection still outperforms linear channel estimation, though by a smaller margin. 
In this journal version of our work, we also include proofs and derivations that were omitted in the conference proceedings due to space limitations.
For sake of readability and convenience to the reader, we repeat parts of the material presented in  \cite{mueller:13a,cottatellucci:13}.

The paper is organized as follows:
In Section~\ref{systemmodel}, we introduce the system model. In Section~\ref{proposedalgorithm}, we introduce the algorithm for nonlinear channel estimation utilizing the array gain. In Section~\ref{performanceanalysis}, we analyze the performance of the algorithm by means of random matrix theory in the limit of large number of base station antennas. 
In Section~\ref{numerics}, we investigate the performance of the algorithm for a finite number of antennas by simulative means.
Finally, Section~\ref{conclusions} concludes the paper.

\section{System Model}
\label{systemmodel}

Consider a wireless communication channel. In order to ease notation and for sake of conciseness, let the channel bandwidth be equal to the coherence bandwidth. Channels whose physical bandwidth is wider than the coherence bandwidth can be decomposed into equivalent parallel narrowband channels by means of orthogonal frequency division multiplexing or related techniques.
The coherence time of the channel measured in symbol intervals is thus given by \cite{rappaport:96}
\begin{equation}
C = \frac{3}{4\sqrt\pi f_0 \tau} \frac cv 
\end{equation}
with $f_0$, $\tau$, $v$, and $c$ denoting carrier frequency, delay spread, mobile speed, and the speed of light, respectively.
Considering extreme values for these parameters, e.g.\ bullet train speed of $v=350$~km/h and very high delay spread $\tau=5$~$\mu$s (which corresponds to an excess distance of 1.5~km) the coherence time at $f_0=2.6$~GHz is 99 symbols. 
For more typical speeds of mobile terminals and/or smaller cells, the coherence time can be one or several orders of magnitude larger.

In the following, we consider the uplink (reverse link) of a cellular massive MIMO system.
Therefore, the number of receive antennas $R$ is much larger than the number of transmit antennas $T$.
The number of transmit antennas is limited by the richness of the propagation channel.
Measurements in Manhattan \cite[Fig.~5]{chizhik:03} show that the richness is limited to around 10 to 13 degrees of freedom.
In more typical outdoor environments with fewer high buildings, the scattering richness will be even lower. 
It is therefore sensible to assume that the number of transmit antennas is small compared to both the number of receive antennas and the coherence time
\begin{equation}
R\gg T \ll C.
\end{equation}
The number of receive antennas is constrained by their physical size. 
At 2.6~GHz, a uniform linear array spaced in half wavelengths reaches the length of $d=6$~m for 104 elements.
Spacing the elements on a two-dimensional grid, the number of receive antennas can be even larger.
Obviously, the number of receive antennas can be larger, equal or smaller than the coherence time.
We will therefore, introduce the normalized coherence time
\begin{equation}
\label{defkappa}
\kappa = \frac CR
\end{equation}
which is assumed to be a finite, non-zero constant throughout this paper.

Let the frequency-flat, block-fading propagation channel  be described by the matrix equation
\begin{equation}
\label{chamod}
\matr Y = \matr{HX} + \matr Z,
\end{equation}
where $\matr X \in \CC^{T\times C}$ is the transmitted data (eventually multiplexed with pilot symbols), 
$\matr H \in\CC^{R\times T}$ is the channel matrix of unknown propagation coefficients, $\matr Y\in\CC^{R\times C}$ is the received signal, and $\matr Z\in\CC^{R\times C}$ is the total impairment.
Furthermore, we assume that channel, data, and impairment have zero mean, i.e.\ $\expect \matr X = \expect \matr H = \expect \matr Z = \matr 0$.
The impairment includes both thermal noise and interference from other cells and is, in general, neither white nor Gaussian.

Note that \eqref{chamod}, can also be understood as a code-division multiple-access (CDMA) system with the columns of $\matr H$ denoting the spreading sequences and $R$ denoting the processing gain.
It is well-known that CDMA can be demodulated without knowledge of the spreading sequences by means of blind algorithms, see e.g.\ \cite{madhow:98}. Many of those algorithms can also be applied in massive MIMO systems.
 In the following section, we introduce an algorithm, which we consider particularly suited for cellular massive MIMO.

\section{Proposed Algorithm}
\label{proposedalgorithm}

\subsection{General Idea}

Before going into the details of the proposed algorithm, we start with the idea behind the proposed procedure.
Consider the channel model \eqref{chamod} for a single active transmit antenna, i.e.\ $T=1$ and look for the matched filter $\matr m^\dagger$ such that the signal-to-noise ratio (SNR) at its output is maximum.
In white noise, maximizing the SNR is equivalent to maximizing the total received power normalized by the power gain of the filter.
Thus, the optimum filter is given by
\begin{equation}
\label{RQ}
\matr m^{\circ} = \argmax\limits_{\matr m} \frac{\matr m^\dagger \matr J \matr m}{\matr m^\dagger\matr m}
\end{equation}
with
\begin{equation}
\matr J = \expect\limits_{\matr X,\matr Z|\matr H} \left\{\matr{YY}^\dagger\right\}.
\end{equation}
It is a well-known result of linear algebra that the vector $\matr m^{\circ}$ maximizing the right hand side of \eqref{RQ}, commonly referred to as the Rayleigh quotient,  is that eigenvector of $\matr J$ that corresponds to the largest eigenvalue of $\matr J$.
Since we do not know the matrix $\matr J$, we have to cope with the approximate solution
\begin{align}
\matr m^\ast &= \argmax\limits_{\matr m} \frac{\matr m^\dagger \matr{YY}^\dagger \matr m}{\matr m^\dagger\matr m}.
\end{align}
This approximation is tight for large number of antenna elements, i.e.\ we have the almost sure convergence of the inner product
\begin{equation}
\label{eq6}
\left|\langle \matr m^\circ ; \matr m^\ast \rangle \right| \to \left|\left| \matr m^\circ\right|\right| \cdot \left|\left| \matr m^\ast\right|\right| 
\end{equation}
as $R\to\infty$, if the largest eigenvalue of the noise is negligible against the largest eigenvalue of the signal, i.e.\
\begin{align}
\label{lb}
\lim\limits_{R\to\infty} \frac{\displaystyle\max\limits_{\matr m} \frac{ \matr m^\dagger \matr{ZZ}^\dagger \matr m}{\matr m^\dagger\matr m}}{\displaystyle \max\limits_{\matr m} \frac{\matr m^\dagger \matr H \matr{XX}^\dagger\matr H^\dagger \matr m}{\matr m^\dagger\matr m}} = 0.
\end{align}
Note that the limit $R\to\infty$ implies $C\to\infty$ due to \eqref{defkappa}.
For finite number of antennas, there are better approximations for $\matr J$ than $\matr{YY}^\dagger$, e.g.\ G-estimation \cite{mestre:08}.
However, such methods exceed the scope of the present work and are left for future research.
G-estimation will further improve the performance of the proposed subspace projection method, in practice.

The limit condition \eqref{lb} is not hard to fulfill. In fact, it holds true for independent constant variance entries in $\matr Z$, $\matr H$, and $\matr X$. 
To see this note that the largest eigenvalue of $\matr{ZZ}^\dagger$ scales linearly with $R$, as the number of entries in $\matr Z\in\CC^{R\times C}$ grows quadratically, but the number of non-zero eigenvalues grows linearly.
At the same time the largest eigenvalue of $\matr H \matr{XX}^\dagger\matr H^\dagger$ grows quadratically with $R$, as the number of entries in $\matr H \in\CC^{R\times T}$ grows linearly, the number of entries in $\matr X\in\CC^{T\times C}$ grows linearly, but the number of non-zero eigenvalues is $T$ and thus constant.  

\subsection{Detailed Algorithm}

Having found an algorithm for a single transmitter and white noise, we now apply this idea to multiple transmit antennas and analyze its performance in colored noise.
Consider the singular value decomposition
\begin{equation}
\label{fullsvd}
\matr Y = \matr{U\Sigma V}^\dagger
\end{equation}
with unitary matrices $\matr U\in\CC^{R\times R}$ and $\matr V\in\CC^{C\times C}$ and the $R\times C$ diagonal matrix $\matr \Sigma$ with diagonal entries $\sigma_1\ge \sigma_2 \ge \cdots \ge \sigma_{\min\{R,C\}}$ sorted in non-increasing order.
As shown in \cite{ngo:12}, the columns of $\matr U$ are highly correlated with the columns of $\matr H$. Based on this observation, \cite{ngo:12} proposes two algorithms for improved nonlinear estimation of the channel matrix $\matr H$.

In the sequel, we pursue a strategy different from the one in \cite{ngo:12}.
We decompose the matrix of left singular vectors
\begin{equation}
\matr U = [ \matr S | \matr N]
\end{equation}
into the signal space basis $\matr S\in\CC^{R\times T}$ and the null space basis $\matr N\in \CC^{R\times(R-T)}$.
Now, we project the received signal onto the signal subspace and get
\begin{equation}
\label{pro}
\matr{\tilde Y} = \matr S^\dagger \matr Y.
\end{equation}
The null space basis $\matr N$ is not required in the sequel. In fact, there is no need to compute the full singular value decomposition \eqref{fullsvd}.
Only the basis of the signal subspace $\matr S$ is needed and there are efficient algorithms available to exclusively calculate $\matr S$.

Consider now the massive MIMO case, i.e.\ $R\gg T$: The $T$-dimensional signal subspace is much smaller than the $R$-dimensional full space, which the noise lives in.
White noise is evenly distributed in all dimensions of the full space. Thus,  the influence of white noise onto the signal subspace becomes negligible as $R\to\infty$. In other words: The considerations for the largest eigenvalue in \eqref{lb} and its corresponding eigenvector in \eqref{eq6} are equally valid for the $T$ largest eigenvalues and their corresponding eigenvectors, as long as $T$ is finite. 

Using the algorithm above, we can achieve an array gain even without the need for estimating the channel coefficients. In fact, channel estimation can be delayed until the received signal has been projected onto the signal subspace and the dominant part of the white noise has already been suppressed.

In order to save complexity it is sensible not to estimate the channel matrix $\matr H$, at all.
Instead, we directly consider the subspace channel
\begin{equation}
\label{projectedchannel}
\matr{\tilde Y} = \matr{\tilde HX} + \matr {\tilde Z}
\end{equation}
and estimate the much smaller subspace channel matrix $\matr{\tilde H} \in\CC^{T\times T}$ by standard methods of linear channel estimation based on pilot symbols.
Hereby, we have transformed the problem of channel estimation for highly asymmetric massive MIMO systems into the well-explored problem of channel estimation for classical symmetric MIMO systems.
Although the data dependent projection \eqref{pro} implies that the noise $\matr {\tilde Z}=\matr S^\dagger\matr Z \in\CC^{T\times C}$ is not independent from the data $\matr X$, neglecting this dependence is an admissible approximation that becomes exact due to \eqref{eq6}, as the number of receive antennas $R$ grows large.

In addition to white noise, there is co-channel interference from $L$ neighboring cells.
For sake of notational convenience, we assume that the number of transmit antennas is identical in all cells and equal to $T$.
The interference from neighboring cells is anything but white. It is the more colored, the smaller the ratio
\begin{equation}
\alpha = \frac{T}R
\end{equation}
which will be called {\em load} in the following.
Any $R$-dimensional channel vector is orthogonal to any other channel vector in the limit $R\to\infty$ \cite{marzetta:10}. This holds regardless whether the two channel vectors correspond to transmitters in the same cell or in different cells. 
In the limit of zero load, i.e.\ $\alpha\to0$, we have an even stronger result: the subspace spanned by the co-channel interference is orthogonal to the signal subspace.\footnote{Note that the pairwise orthogonality of channel vectors holds for $R\to\infty$, in general, and does not require $\alpha\to0$. However, the orthogonality of subspaces requires $\alpha\to0$ in addition to $R\to\infty$, as the accumulation of $T=\alpha R$ vanishing pairwise correlations is not vanishing, in general.}
That means that in the limit $R/T\to\infty$, the $(L+1)T$ largest singular values of the received signal matrix $\matr Y$ become identical to the Euclidean norms of the $(L+1)T$ channel vectors.
We only need to identify which singular values correspond to channel vectors from inside the cell as opposed to channel vectors from transmitters in neighboring cells. Then, we can remove the interference from neighboring cells by subspace projection.

\subsection{Identifying Signals of Interest}
\label{isi}
Note that for $R\to\infty$, the system has infinite diversity and the effect of short-term fading (Rayleigh fading) vanishes.
Thus, the norm of a channel vector is solely determined by path loss and long-term fading (shadowing).
In a cellular system with perfect received power control and a power-controlled handoff strategy, the norm of channel vectors from neighboring cells can never be greater than the norm of channel vectors from the cell of interest.
We conclude that the identification of singular values belonging to transmitters within the cell of interest is possible by means of ordering them by magnitude in the limit $(R,\alpha) \to (\infty,0)$, i.e.\ the number of receive antennas grows large while the number of transmit antennas does not.

For practical systems with small, but nonzero load, i.e.\ $0<\alpha\ll 1$, a certain power margin is required between signals of interest and interfering signals. For most interfering users, such a power margin is created for free by shadowing and path loss. However, there might be few users close to cell boundaries who lack such a power margin.
As a kind of countermeasure, a power margin has to be engineered for them. There are various ways to do so. In the sequel, we will exemplarily list two such potential methods.

One way to create an additional power margin is a smart choice of frequency or time re-use patterns. However, this requires coordination among cells. Another way to create an additional power margin is to equip each user with at least two transmit antennas.
Then, the few users who suffer from insufficient power margin can form beams that favor one of the base stations or access points over others\footnote{Note that such beam forming does not require channel state information. One can keep on forming random beams until a sufficient power margin is reached.}. This will noticeable increase their power margins. The majority of users will not need to employ such methods and can use the two antennas for spatial multiplexing.

\section{Performance Analysis}
\label{performanceanalysis}

We have demonstrated above, that the proposed algorithm works in principle in massive MIMO systems if the number of receive antennas is much larger than the product of transmit antennas and neighboring cells. In practical systems, the number of transmit and receive antennas is finite and the load $\alpha$ can be made very small but not arbitrarily small.
The standard assumption of massive MIMO systems, i.e.\ $R\to\infty$, while $T$ staying finite, gives overoptimistic results for finite systems of practical interest.
In order to find the limits of the subspace projection method, we need to consider a more refined limit in the number of antennas.
For that purpose, we utilized the asymptotic scale invariance of the eigenvalue spectra of large random matrices: For appropriate normalization of the entries, the spectra of random matrices are hardly affected (asymptotically invariant) if all matrix dimensions scale proportionally.
Thus, a useful and insightful approach to understand the behavior of a real cellular massive MIMO network consists in assuming that both $T$ and $R$ grow large with a small, but fixed ratio $\alpha.$ 
This approach is tantamount to studying a system with $T=5n$ transmit antennas, $R=300n$ receive antennas and coherence time $C= 100n$ for $n\to\infty$ and assume  that the result for a system with $T=5$, $R=300$ and $C=100$ hardly differs.


We decompose the impairment process
\begin{equation}
\label{noisedecomp}
\matr Z = \matr W + \matr H_{\rm I} \matr {X}_{\rm I}
\end{equation}
into white noise $\matr W$ and interference from $L$ neighboring cells where interfering data $\matr X_{\rm I}\in \CC^{LT\times R}$ is transmitted in neighboring cells and received in the cell of interest through the channel $\matr H_{\rm I}\in\CC^{R\times LT}$.
Combining \eqref{chamod} and \eqref{noisedecomp}, we get
\begin{equation}
\label{newmodel}
\matr Y = \matr {HX} + \matr H_{\rm I}\matr X_{\rm I} + \matr W.
\end{equation}

Let the entries of the data signal $\matr X$ be iid with zero mean and variance $P$. Let the entries of the channel matrix
$\matr H$ be also iid with zero mean, but have unit variance. 
Let the entries of the matrix of interfering signals $\matr X_{\rm I} $ be iid with zero mean and variance $P$ and let the entries of the $k^{\rm th}$ column of the matrix of interfering channels $\matr H_{\rm I}$ be independent with zero mean and variance $I_k/P$ such that the ratio $I_k/P$ accounts for the relative attenuation between out-of-cell user $k$ and the intracell users.
Let the empirical distribution of $I_k$ converge to a limit distribution as $LT\to\infty$ which is denoted by ${\rm P}_I(\cdot)$.
Furthermore, we assume that the elements of the noise $\matr W$ are independent and identically distributed (iid) with zero-mean and variance $W$.
Let us denote the asymptotic eigenvalue distribution of $\matr {YY}^\dagger$ as ${\rm P}_{\matr {YY}^\dagger}(x)$.
In Appendix \ref{app:stieltjes_transform}, we show that this asymptotic eigenvalue distribution obeys
\begin{align}
s{\rm G}_{\matr
{YY}^\dagger} \left(s\right)+1 =&-  \frac {PTC \alpha \left(s{\rm G}_{\matr {YY}^\dagger}\left(s\right)+{1-\kappa}{}\right){\rm G}_{\matr
  {YY}^\dagger}(s)}{\alpha\kappa-PTC
 \left(s{\rm G}_{\matr {YY}^\dagger}\left(s\right)+{1-\kappa}{}\right) {\rm G}_{\matr {YY}^\dagger}(s)}\nonumber\\
 &-\int
 \frac {xLTC \alpha \left(s{\rm G}_{\matr {YY}^\dagger}\left(s\right)+{1-\kappa}{}\right){\rm G}_{\matr {YY}^\dagger}(s){\rm dP}_I(x)}{\alpha\kappa
 -xTC
 \left(s{\rm G}_{\matr {YY}^\dagger}\left(s\right)+{1-\kappa}{}\right){\rm G}_{\matr {YY}^\dagger}(s)}\nonumber\\
 &-
 \frac {WC  \left(s{\rm G}_{\matr {YY}^\dagger}\left(s\right)+{1-\kappa}\right){{\rm G}_{\matr {YY}^\dagger}(s)}}{\kappa}
\label{fixedpoint}
\end{align}
with
\begin{equation}
{\rm G}_{\matr {YY}^\dagger}(s) = \int \frac{{\rm dP}_{\matr {YY}^\dagger}(x)}{x-s}
\end{equation}
denoting its Stieltjes transform.
By means of the Stieltjes inversion formula
\begin{equation}\label{eq:stieltjes_inverse}
{\rm p}(x) = \frac 1\pi \lim\limits_{y\to0^{+}} \Im {\rm G}(x+{\rm j} y)
\end{equation}
the asymptotic eigenvalue density is obtained. 

\begin{figure}
\centerline{\epsfig{file=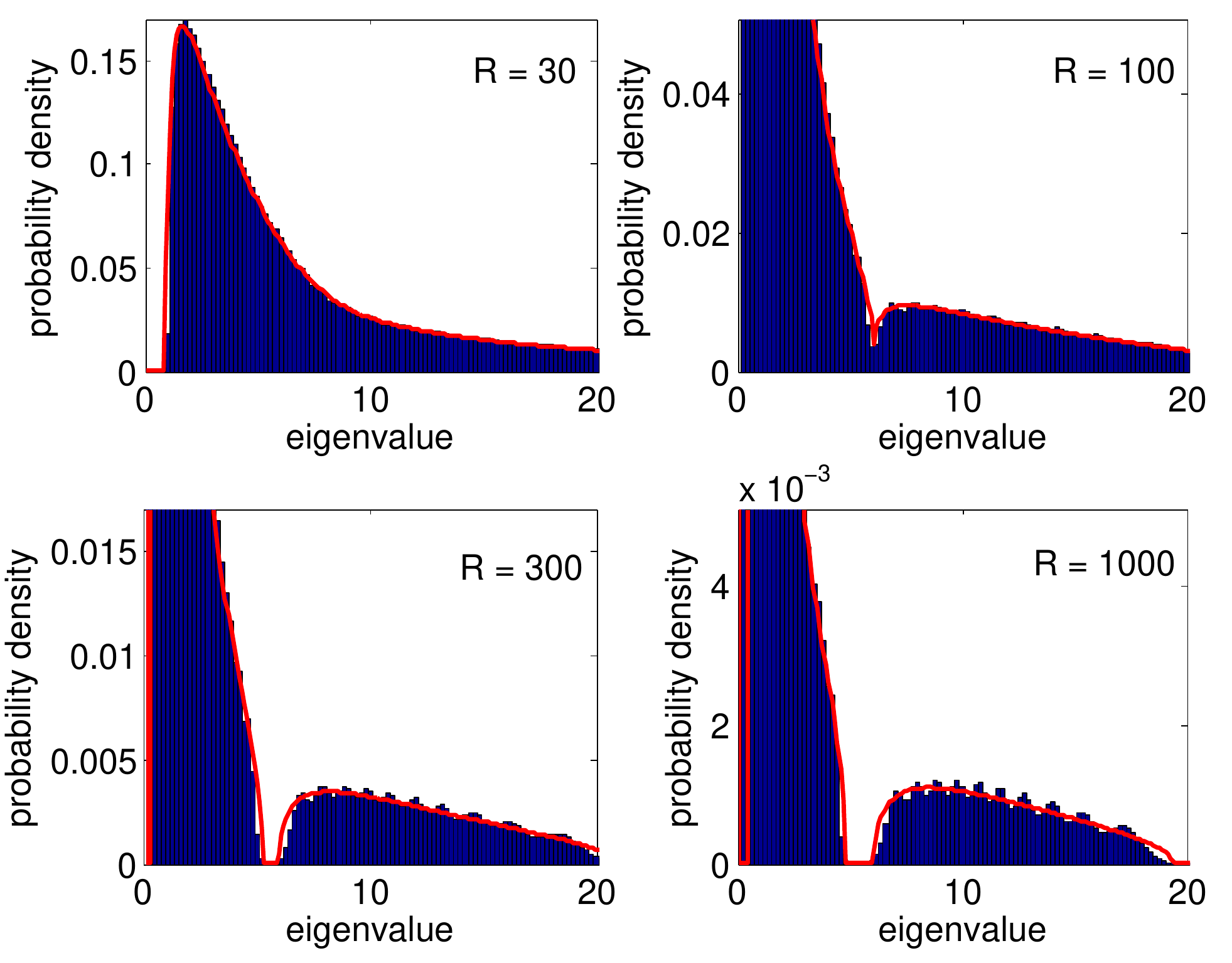,width=.8\columnwidth}}
\caption{\label{fig_asymptotic_evd} Asymptotic eigenvalue density of the matrix $\matr {YY}^\dagger/R$ in solid red line for $\frac\alpha\kappa=\frac1{10}$, $P=-10$ dB, $I_k=\frac{P(k {\rm mod} T)}{4T}\, \forall k=1\dots LT$, $L=2$, $W=0$ dB. The asymptotic eigenvalue distribution is compared to the empirical eigenvalue density for $T=10,$ $C=100$, and various value of $R$ given by the histograms in blue. }
\end{figure}

In Figure \ref{fig_asymptotic_evd}, the solid lines in red shows the asymptotic eigenvalue distribution of $\matr{YY}^\dagger/R$ obtained by (\ref{fixedpoint})-(\ref{eq:stieltjes_inverse}).
The histograms in blue show the empirical eigenvalue distributions of $\matr {YY}^\dagger/R$ for $T=10$, $C=100$, and various values of $R$. 
We observe that for sufficiently large number of receive antennas $R$, the distribution decomposes into two disjunct bulks: A noise and interference bulk to the left and a bulk of the signal of interest to the right.
If the bulks do not overlap, we can blindly separate the signals of interest from interference and noise as discussed in Section~\ref{isi}. 

The bulks are not disjunct in general, but only for certain values of the involved system parameters. 
It is therefore of utmost importance for practical design of blind pilot decontamination to know which system parameters do lead to bulk separation. 
The extremely good match between the asymptotic distribution and the empirical distribution for finite matrices corroborate the usefulness to study the support of the asymptotic eigenvalue distribution of $\matr{YY}^\dagger$ and the asymptotic conditions of bulk separability.
We remark that, in principle, the common interference and noise bulk could also separate into two separate bulks for noise and interference. However, this happens only if the weakest interfering signal is sufficiently strong in comparison to the noise power.

\subsection{Unilateral Approximation}
\label{appana}

The general result for the asymptotic eigenvalue distribution \eqref{fixedpoint} is implicit and not very intuitive. In the following, we develop an approximate analysis for small, but not vanishing loads $\alpha$. It is based upon the separate calculation of each bulk and subsequent rescaling of the bulks due to pairwise bulk-to-bulk repulsion.
We will see that it leads to explicit and intuitive  design guidelines.

In the large antenna limit $R=C/\kappa \to\infty$, the singular values of $\matr W/\sqrt{CW}$ follow the Marchenko-Pastur law, i.e.\
\begin{equation} \prob{\matr W\,}x = \frac{\sqrt{\frac4\kappa-(x-1-\frac1\kappa)^2}}{\pi x} \end{equation} for $1/\sqrt\kappa -1
<x<1/\sqrt\kappa +1$. In the worst case, the $T$ largest singular values of the noise affect the signal of interest. The power of white
noise being present in $\matr{\tilde Y}$ is thus at most \begin{equation}\label{eq:max_noise_bulk_alpha0} T C W \left(1+\frac1{\sqrt\kappa}\right)^2. \end{equation}

The total power of the signal of interest at the receiver is $TRCP$
and the signal-to-noise ratio in $\matr {\tilde Y}$ is lower bounded by \begin{equation} {\rm SNR} \ge \frac PW
\frac{R}{\left(1+\frac1{\sqrt\kappa}\right)^2} \ge \frac PW \cdot \frac {\min\{R,C\}}4
\end{equation} 
For fixed normalized coherence time, the first lower bound on the signal-to-noise ratio scales linearly with the number of receive
antennas $R$.
For fixed absolute coherence time, second lower bound scales with the minimum of the coherence time and the number of receive antennas.
Note, however, that the two lower bounds can be quite loose and the actual SNR might be considerably larger.

In addition to white noise, there is co-channel interference from neighboring cells.
The co-channel interference is not white but, like the signal of interest, highly concentrated in certain subspaces.
The empirical distribution of the squared singular values of the normalized signal of interest, i.e.\ $\matr{HX}/\sqrt{TR}$, is shown in \cite{mueller:13a} to converge, as $R\to\infty$, to a limit distribution which for $\alpha \ll1$ is supported in the interval
\begin{equation}
\label{Pint}
{\cal P} = \left[\frac {\kappa P}\alpha- 2P\sqrt{\frac{\kappa^2+\kappa}\alpha}; \frac{\kappa P}\alpha +  2P\sqrt{\frac{\kappa^2+\kappa}\alpha}\right].
\end{equation}
The empirical distribution of the squared singular values of the normalized co-channel interference, i.e.\ $\matr H_{\rm I}\matr X_{\rm I}/\sqrt{TR}$, also converges to a limit distribution. For $\alpha\ll1$, it is supported in the interval
\begin{equation}
\label{Iint}
{\cal I} = \left[\frac {\kappa I}{\alpha}- 2I\sqrt{L\, \frac{\kappa^2+\kappa}{\alpha}}; \frac{\kappa I}{\alpha} +  2I\sqrt{L\, \frac{\kappa^2+\kappa}{\alpha }}\right]
\end{equation}
for $I_k=I\,\forall k$.
We remark that the condition $I_k=I\,\forall k$ is unrealistic, in practice. However, the general case is not tractable by analytic means.
We note, however, that setting all interference powers to the maximum interference power among the users is a worst case scenario.

When separately calculating the eigenvalue spectra of the signal-of-interest, the interference and the noise, the accuracy of the results suffers from the eigenvalues in different bulks repelling each other. In the following, we will correct for this effect up to first order. We decompose one bulk of eigenvalues into single eigenvalues. Then, we introduce correction factors that account for the scaling of one of the single eigenvalues due to the presence of one other bulk of eigenvalues. We will then approximate the influence of several other bulks, e.g.\ noise bulk and interference bulk, by multiplying the correction factors. This procedure is an approximation, since we neglect the fact that also the scaled bulk of eigenvalues repels the scaling bulk and that the two scaling bulks repel each other.\footnote{To better understand this procedure consider a system with sun (noise bulk), earth (interference bulk) and moon (bulk of signal of interest) which mutually affect each other by gravity. We decompose the moon into single atoms. These atoms are too small to affect the path of the earth. So we can calculate the position of the moon atoms without accounting for the force the moon enacts onto the earth. Then, we apply the same procedure for the interaction between moon atoms and the sun and superimpose the results of the earth-moon and sun-moon interactions. The fact, that sun and earth influence each other is also ignored. }

The presence of additive noise scales the eigenvalues of both the signal of interest and the interference. As shown in Appendix~\ref{app:shift}, the scale factors are given for $R\gg T$ by
\begin{equation}
\label{ns1}
n_{\rm P}= \left(1+\frac{W}{P R}\right)\left(1+\frac{W}{PC}\right)
\end{equation}
and
\begin{equation}
\label{ns2}
n_{\rm I}= \left(1+\frac{W}{IR}\right)\left(1+\frac{W}{IC}\right),
\end{equation}
respectively. Note that the two scale factors converge to 1 in the large system limit irrespective of the load $\alpha$, if the noise power $W$ does not scale with the system size.

The presence of interference scales the eigenvalues of the signal of interest and vice versa. As shown in Appendix~\ref{app:shift}, the scale factors for non-overlapping bulks are given for $R\gg T$ by 
\begin{equation}
i_{\rm P}= \left(1+\frac{L\alpha/\kappa}{\frac PI-1}\right)\left(1+\frac{L\alpha}{\frac PI-1}\right)
\end{equation}
and
\begin{equation}
i_{\rm I}= \left(1+\frac{\alpha/\kappa}{\frac IP-1}\right)\left(1+\frac{\alpha}{\frac IP-1}\right)
,
\end{equation}
respectively. Note, however, that these scale factors are only accurate if $P\gg I$. 
This somehow limits their usefulness, in practice.

If the two supporting intervals do not overlap, i.e.\
\begin{equation}
n_{\rm P} i_{\rm P} {\cal P} \cap n_{\rm I} i_{\rm I} {\cal I} = \emptyset
\end{equation}
or equivalently
\begin{eqnarray}
\label{RMTthreshold}
\frac PI  &> & \frac{n_{\rm I}i_{\rm I}}{n_{\rm P}i_{\rm P}} \cdot \frac{1+2\sqrt{\alpha L \left(1+\frac1\kappa\right)}}{1-2\sqrt{\alpha  \left(1+\frac1\kappa\right)}},
\end{eqnarray}
the singular value distribution of the sum of the signal of interest and the interference converges, as $R\to\infty$, to a limit distribution that is composed of two separate non-overlapping bulks \cite{mueller:12}.
Note that in the limit $\alpha\to0$, the signal bulk always separates from the interference bulk as long as $P/I>1$.
Therefore, the signal subspace and the interference subspace can be identified blindly.
The interference can be nulled out and pilot contamination does {\em not} happen.

\subsection{Bilateral Approximation at High SNR}
The previous approximation was intuitive, but its accuracy is limited. In this subsection, we use perturbation theory for a more precise approximation for small load $\alpha\ll 1$ and $I_k=I$, where we account for the mutual interaction between the interference bulk and the bulk of the signal of interest.

Let us denote by $\mathcal{P}_{W}$ and $\mathcal{I}_{W}$  the eigenvalue bulks corresponding to the signal  subspace and the interference subspaces, respectively, when the white noise variance is $W$. Additionally, let us assume that $P>I$ as in systems of practical interest. Finally, let us define
\begin{align}
r&= \frac \alpha{PTC}=\frac1{PRC} \label{def:r}\\
t&= \frac \alpha{ITC}=\frac1{IRC} \label{def:t}\\
\zeta&=WC. \label{def:zeta}
\end{align}
The following results are shown in Appendix \ref{app:noiseless_system_finite_alpha}: In the high SNR regime, i.e. for $W=0,$ the inverse of the Stieltjes transform is well approximated by the rational function
\begin{equation}\label{eq:inverse_stieltjes_1_order_appr}
    s^{(1)}(G)=\frac{((L+1)(\kappa-2)\alpha -\kappa)G^2 +((Lr+t)(\kappa-1)\alpha -\kappa(r+t))G - \kappa r t}{G((\kappa+2(L+1) \alpha)G^2+((Lr+t)\alpha+\kappa(r+t))G+\kappa rt)}.
\end{equation}
The extremes of the function $s^{(1)}(G)$ are the solutions $G_1,G_2, G_3, G_4$ to the quartic equation \begin{multline}\label{eq:quartic_eq_extremes}
    (2(L+1)^2(\kappa-2)\alpha^2+(L+1)(\kappa-4)\kappa \alpha-\kappa^2)G^4\\+(2((2(Lr+t))(L+1)(\kappa-1)\alpha^2+((Lr+t)(\kappa-1)-(2(L+1))(t+r))\alpha \kappa-(t+r)\kappa^2))G^3\\+((Lr+t)^2(\kappa-1)\alpha^2+(t^2+L r^2)(\kappa-2)\kappa \alpha-(6(L+1))r t \kappa \alpha-((t+r)^2+2rt)\kappa^2)G^2 \\ -2rt\kappa((Lr+t)\alpha+(t+r)\kappa)G-\kappa^2 r^2 t^2=0.
\end{multline}
If $G_i,$ for $i=1,2,3,4,$ are all real with $G_1<G_2<G_3<G_4$ and $s^{(1)}(G_2)<s^{(1)}(G_3)$   then an approximation of  $\mathcal{P}_0$ and $\mathcal{I}_{0}$ is given by
\begin{align}\label{eq:tight_interval}
    \mathcal{P}_0&\approx[s^{(1)}(G_3), s^{(1)}(G_4)] \\
    \mathcal{I}_0&\approx[s^{(1)}(G_1), s^{(1)}(G_2)] \nonumber
\end{align}
and the two intervals $[s^{(1)}(G_3), s^{(1)}(G_4)]$ and $[s^{(1)}(G_1), s^{(1)}(G_2)]$ are disjoint.

\begin{figure}
\epsfig{file=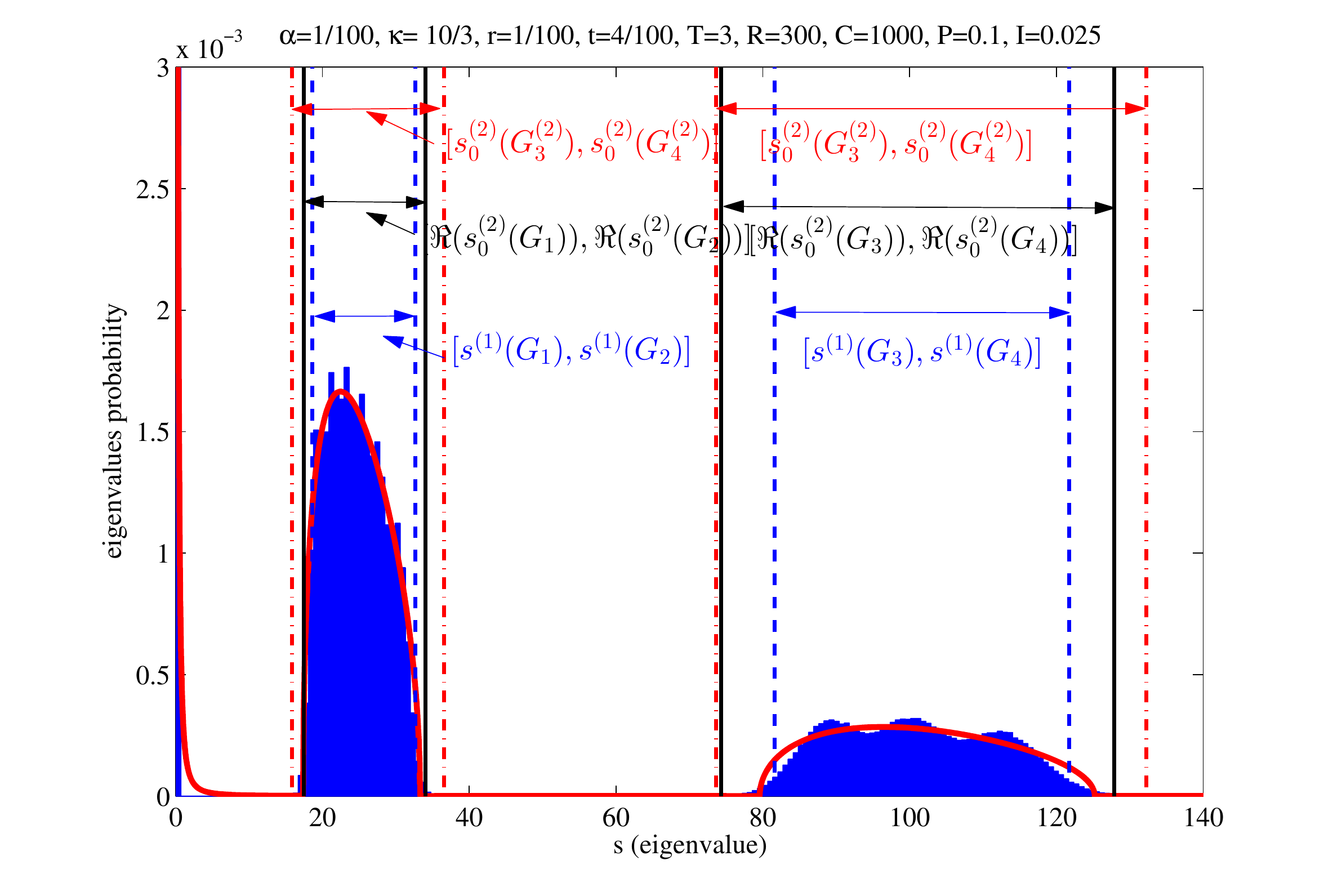,width=\columnwidth}
\caption{\label{fig:tight_interval}  Bulk-support approximation of the asymptotic eigenvalue density of the matrix $\matr {YY}^\dagger/R$ by (\ref{eq:tight_interval}) and (\ref{eq:second_order_tight_interval}) for $\alpha=\frac{1}{100},$ $\kappa=\frac{10}{3},$ $P=-10$ dB, $I_k=\frac{P}{4}\,\forall k$. A further approximation $[\Re(s^{(2)}_0(G_1)), \, \Re(s^{(2)}_0(G_2))]\bigcup [\Re(s^{(2)}_0(G_3)), \, \Re(s^{(2)}_0(G_4))]$ is also shown. The histogram in blue is the empirical eigenvalue density for $T=3,$ $R=300,$ and $C=1000$ while the red line is the asymptotic eigenvalue density. }
\end{figure}

Different approximations of $\mathcal{P}_0$ and $\mathcal{I}_0$ can be obtained by approximating $s(G)$ by the function
\begin{equation}\label{eq:inverse_stieltjes_2_order_appr}
   s_0^{(2)}(G)=\left\{
     \begin{array}{ll}
       \phi_{0}(G)+\rho_{0}(G), & G \in [G_{-}^{\infty}, \, G_{+}^{\infty}] \\
       \phi_{0}(G)-\rho_{0}(G), & \hbox{elsewhere.}
     \end{array}
   \right.
\end{equation}
where $G_{-}^{\infty}$ and $G_{+}^{\infty}$  are the instances of
 \begin{align}\label{eq:poles_s_alpha_1}
   G^{\infty} = &\frac{\kappa(r+t)+\alpha(t+Lr)}{-2\kappa-4\alpha-4L\alpha}\pm 
    \frac {\sqrt{\kappa^2(r-t)^2+2\alpha \kappa (L r^2+ t^2-3 rt-3L t r)+\alpha^2 (t+L  r)^2}}{-2\kappa-4\alpha-4L\alpha}
\end{align}
 with minus and plus sign, respectively,
\begin{align}
    \phi_{0}(G)=\, &\frac{(2\alpha(L+1)(\kappa-1)+\kappa(\kappa-4))G^2+ \kappa(\alpha(t+Lr)+(\kappa-2)(t+r))G+\kappa^2 rt}{{2G^2\left((2\kappa+(L+1)\alpha)G+\kappa(t+r)\right)}},
    \label{def:phi_0} 
    \end{align}\begin{align}
    \rho_{0}(G)=\,& \kappa\left[\kappa(\kappa-4\alpha(L+1))G^4+ 2\kappa(\kappa(t+r) \right.-3\alpha(Lr+t))G^3 +((t^2 +4rt+r^2)\kappa^2\nonumber \\
    & -2\alpha\kappa(Lr-t)(r-t)+\alpha^2(t+Lr)^2)G^2 \left .2\kappa rt(\kappa(t+r)+\alpha(t+Lr))G + \kappa^2t^2r^2 \right]^{1/2}\nonumber \\
    & \quad \times \frac1{2G^2\left((2\kappa+(L+1)\alpha)G+\kappa(t+r)\right)}. \label{def:rho_0}
\end{align}
This approximation of the inverse Stieltjes transform is derived in Appendix \ref{app:noiseless_system_finite_alpha}. The extremes of this function cannot be derived in close form. Then, we approximate them by the zeros of $\rho_0(G)$, $G_1^{(2)},$ $G^{(2)}_2,$ $G^{(2)}_3,$ and $G^{(2)}_4.$ If $G^{(2)}_1<G^{(2)}_2<G^{(2)}_3<G^{(2)}_4$
and $s_0^{(2)}(G^{(2)}_2)<s_0^{(2)}(G^{(2)}_3),$ we obtain the approximations
\begin{align}\label{eq:second_order_tight_interval}
    \mathcal{P}_0&\approx[s^{(2)}_0(G^{(2)}_3), s^{(2)}_0(G^{(2)}_4)]=[\phi_0(G^{(2)}_3), \phi_0(G^{(2)}_4)] \\
    \mathcal{I}_0&\approx[s^{(2)}_0(G^{(2)}_1), s^{(2)}_0(G^{(2)}_2)]=[\phi_0(G^{(2)}_1), \phi_0(G^{(2)}_2)]. \nonumber
\end{align}
which are motivated in Appendix \ref{app:noiseless_system_finite_alpha}.
The approximated intervals in (\ref{eq:tight_interval}) and (\ref{eq:second_order_tight_interval}) obtained by application of perturbation theory are a very good approximation of $\mathcal{P}_0$ and $\mathcal{I}_0$ as shown in Figure \ref{fig:tight_interval}.  The approximation obtained by (\ref{eq:second_order_tight_interval}) contains the support of the asymptotic eigenvalue distribution. 

As well known, the quartic equations to determine $G_i$ and $G_i^{(2)},$ $ i=1, \ldots 4,$ admit solutions in closed form. However, they are not insightful and handy because of their complexity. Thus, in the following, we propose looser approximations of the intervals $\mathcal{P}_0$ and $\mathcal{I}_0$ yielding handier conditions on bulk separation. Further approximations yield

\begin{equation}\label{eq:approx_P0}
    \mathcal{P}_{0}\subset \left[ s^{(2)}_{\mathcal{P}}(G_{\mathcal{P}_{\ell}}), \, s_{\mathcal{P}}^{(2)}(G_{\mathcal{P}_{u}}) \right]
\end{equation}
where
\begin{align}\label{eq:s_P^(2)}
s_{\mathcal{P}}^{(2)}(x)&= \frac{2\alpha\kappa(L+1)-2\alpha(L+1)+2\kappa(1-\kappa)}{2((1+L)\alpha-\kappa)x+2\kappa(t-2r)}
+\frac{\kappa(\kappa(t-5r)+\alpha(t+Lr)+4r-2t)x+\kappa^2r(t-3r)}{2x^2(((1+L)\alpha-\kappa)x+\kappa(t-2r))}
\end{align}
and $G_{\mathcal{P}_{\ell}}$ and $G_{\mathcal{P}_{u}}$ equal the instances of 
\begin{equation}\label{eq:approx_estremes_P}
    G_{\mathcal{P}}^{(2)}=-\kappa r(t-r)\frac{\kappa(t-r)+\alpha(t+(L-2)r)\pm 2\sqrt{\alpha \kappa (t-r)^2-\alpha^2r(t+(L-1)r)}}{(\alpha t+\alpha Lr-\kappa t+\kappa r)^2+4 \alpha \kappa L r(t-r)}
\end{equation}
which is obtained by selecting the plus and minus sign, respectively. Similarly, for the bulk associated to the interference subspace,
\begin{equation}\label{eq:approx_I0}
    \mathcal{I}_{0}\subset \left[ s_{\mathcal{I}}^{(2)}(G_{\mathcal{I}_{\ell}}), \, s_{\mathcal{I}}^{(2)}(G_{\mathcal{I}_{u}}) \right]
\end{equation}
where
\begin{align}\label{eq:s_I^(2)}
s_{\mathcal{I}}^{(2)}(x)&= \frac{2\alpha\kappa (L+1) -2\kappa(\kappa-1)-2\alpha(L+1)}{2((\alpha(L+1)-\kappa)x-\kappa(2t-r))} +\frac{\kappa((4-5\kappa)t+(\kappa-2) r+ \alpha ( t+Lr))x+\kappa^2 t (r-3t)}{2x^2((\alpha(L+1)-\kappa)x-\kappa(2t-r))}
\end{align}
and $G_{\mathcal{I}_{\ell}}$ and $G_{\mathcal{I}_{u}}$ are obtained by selecting the instance of 
\begin{equation}\label{eq:approx_estremes_I}
    G_{\mathcal{I}}=-\kappa t(t-r)\frac{\kappa(t-r)+\alpha((2L-1)t-Lr)\pm 2\sqrt{\alpha \kappa L (t-r)^2+\alpha^2Lt((L-1)t-Lr)}}{(\alpha t+\alpha Lr-\kappa t+\kappa r)^2+4 \alpha \kappa L r(t-r)}
\end{equation}
with plus and minus sign, respectively. 
The derivation of the proposed approximations for $\mathcal{P}_0$ and $\mathcal{I}_0$ is detailed in Appendix \ref{app:noiseless_system_finite_alpha}.

By enforcing $G_{\mathcal{I}_u} < G_{\mathcal{P}_{\ell}} $ we obtain a bound on the ratio $\frac{\alpha}{\kappa}$ as a decreasing function of the ratio $\beta=\frac{r}{t}=\frac{I}{P}$
\begin{equation}\label{eq:separability_region_boundary}
    \frac{\alpha}{\kappa} \leq \frac{(1-\beta)^2 (L \beta^2+3(L+1)\beta+1-2(1+\beta)\sqrt{3L \beta} )}{(L\beta^2-1)(L \beta^2+6(L-1)\beta-1)+(9L^2-2L+9)\beta^2}.
\end{equation}
\begin{figure}
\centerline{\epsfig{file=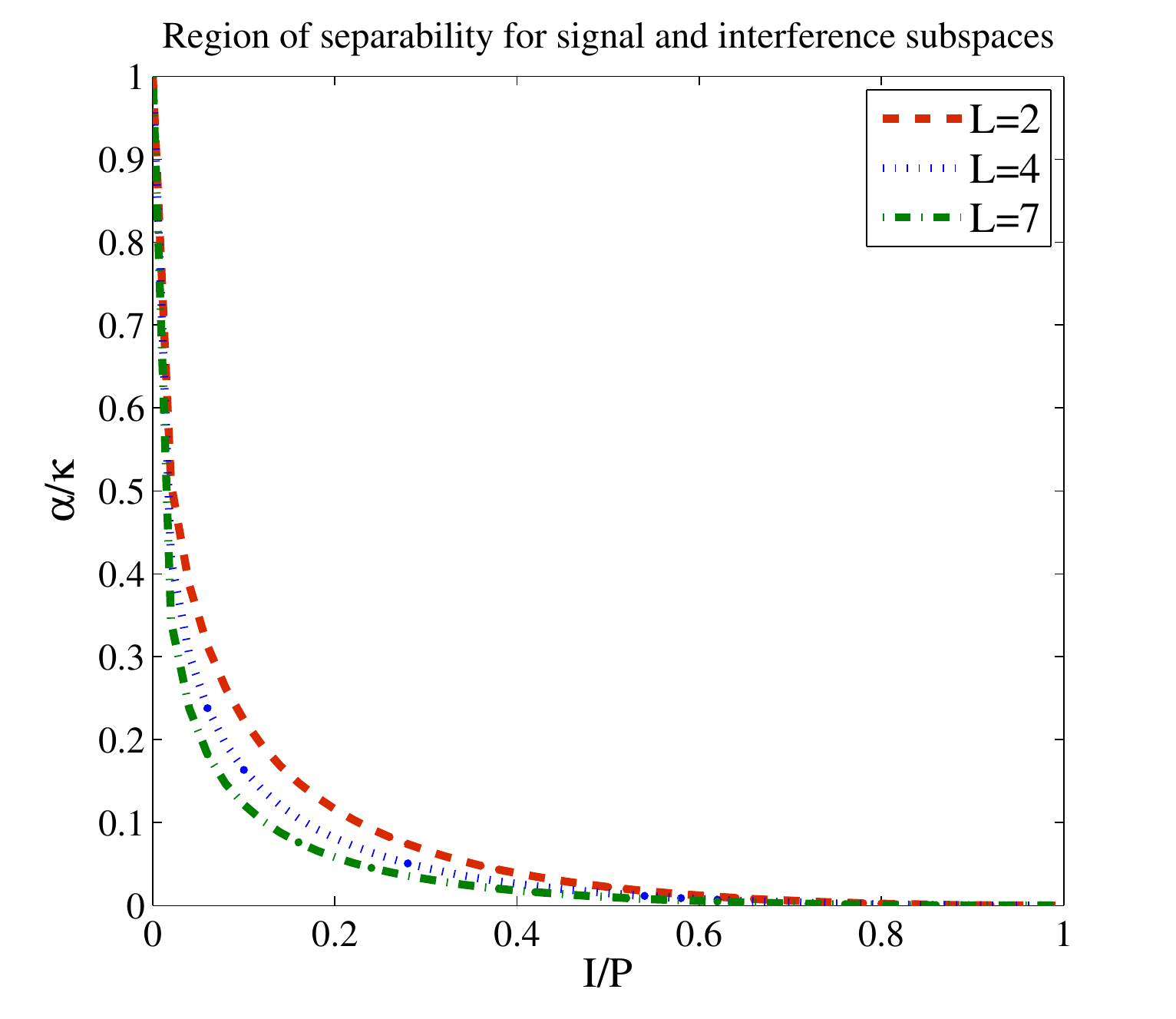,width=.7\columnwidth}}
\caption{\label{fig:separability_region} Separability region obtained by (\ref{eq:separability_region_boundary}) for $L=2,4,$ and 7. }
\end{figure}
Figure \ref{fig:separability_region} shows the region of parameters $\frac{\alpha}{\kappa}$ and $\beta$ where the bulks of the eigenvalues for  the signals of interest and the one for interference do not overlap for various values of $L.$ As expected, the separability region shrinks when the interference from adjacent cells increases, i.e., $L$ increases.

\subsection{Bilateral Approximation for General SNR}

An approach similar to the one proposed for the high SNR regime in Appendix \ref{app:noiseless_system_finite_alpha} can be applied for $W >0$ to determine  approximated supports of the bulks $\mathcal{P}_W$ and $\mathcal{I}_W.$ We propose the conclusive results in the following while the derivations are detailed in Appendix \ref{app:noisy_system_bulk_boundary}.
Then,
\begin{equation}\label{eq:approx_PW}
    \mathcal{P}_{W}\subset \left[ \varsigma_{\mathcal{P}}^{(2)}(\Gamma_{\mathcal{P}_{\ell}}), \, \varsigma_{\mathcal{P}}^{(2)}(\Gamma_{\mathcal{P}_{u}}) \right]
\end{equation}
where 
\begin{align}\label{eq:varsigma_P}
\varsigma_{\mathcal{P}}^{(2)}(x)=& \frac{\alpha\kappa(L+1)-\alpha(L+1)+\kappa(1-\kappa)+\zeta (\kappa-1)(t-2r)}{((1+L)\alpha-\kappa+\zeta(t-2r))x+\kappa(t-2r)}
\nonumber \\
&+\frac{\kappa}{2x^2} \frac{(\kappa(t-5r)+\alpha(t+Lr)+4r-2t+\zeta r (t-3r))x+\kappa r(t-3r)}{((1+L)\alpha-\kappa+\zeta(t-2r))x+\kappa(t-2r)}
\end{align}
and $\Gamma_{\mathcal{P}_{\ell}}$ and $\Gamma_{\mathcal{P}_{u}}$ equal the instances of 
\begin{equation}\label{eq:Gamma_P}
    \Gamma_{\mathcal{P}}^{(2)}=\frac{-\kappa r(t-r)\left[\zeta r(t-r)+\kappa(t-r)+\alpha(t+(L-2)r)\pm 2\sqrt{\alpha \kappa (t-r)^2-\alpha^2r(t+(L-1)r)}\right]}{(r(t-r)\zeta+(\alpha-\kappa)t+(\alpha L+\kappa)r)^2+4r\zeta ((\alpha+\kappa)r^2-(\alpha+2\kappa)tr+\kappa t^2)+4\alpha \kappa L r(t-r)}
\end{equation}
which are obtained by selecting the plus and minus sign, respectively. Similarly, for the bulk associated to the interference subspace,
\begin{equation}\label{eq:approx_IW}
    \mathcal{I}_{W}\subset \left[ \varsigma_{\mathcal{I}}^{(2)}(\Gamma_{\mathcal{I}_{\ell}}), \, \varsigma_{\mathcal{I}}^{(2)}(\Gamma_{\mathcal{I}_{u}}) \right]
\end{equation}
where 
\begin{align}
\varsigma_{\mathcal{I}}^{(2)}(x)=& \frac{\alpha\kappa (L+1) -\kappa(\kappa-1)-\alpha(L+1)-(\kappa-1)(2t-r)\zeta}{(\alpha(L+1)-\kappa-\zeta(2t-r))x-\kappa(2t-r)} \nonumber \\ & +\frac{\kappa((4-5\kappa)t+(\kappa-2) r+ \alpha ( t+Lr)-\kappa \zeta t (3t-r))x+\kappa^2 t (r-3t)}{2x^2((\alpha(L+1)-\kappa-\zeta(2t-r))x-\kappa(2t-r))} \label{eq:varsigma_I}
\end{align}
and $\Gamma_{\mathcal{I}_{\ell}}$ and $\Gamma_{\mathcal{I}_{u}}$ are obtained by selecting the instance of 
\begin{equation}\label{eq:Gamma_I}
    \Gamma_{\mathcal{I}}=\frac{-\kappa t(t-r)\left[\kappa(t-r)+\alpha(2L-1)t-\alpha Lr+t(t-r)\zeta \pm 2\sqrt{\alpha \kappa L (t-r)^2+\alpha^2Lt((L-1)t-Lr)}\right]}{(\alpha t+L \alpha r-t \kappa+r \kappa+ t(t-r) \zeta)^2+4 (t-r)( t[(\kappa+\alpha L-\alpha) t-(\alpha L+\kappa) r]\zeta+\alpha \kappa L r)}
\end{equation}
with plus and minus sign, respectively.

Interestingly, the separability condition obtained by enforcing $\Gamma_{\mathcal{I}_u} < \Gamma_{\mathcal{P}_{\ell}} $ yields to condition (\ref{eq:separability_region_boundary}) as in the case of absence of noise.
This is not as surprising as it may look at first sight, as it was already observed from \eqref{ns1} and \eqref{ns2} that the noise does not affect the support in the large system limit.

\begin{figure}
\epsfig{file=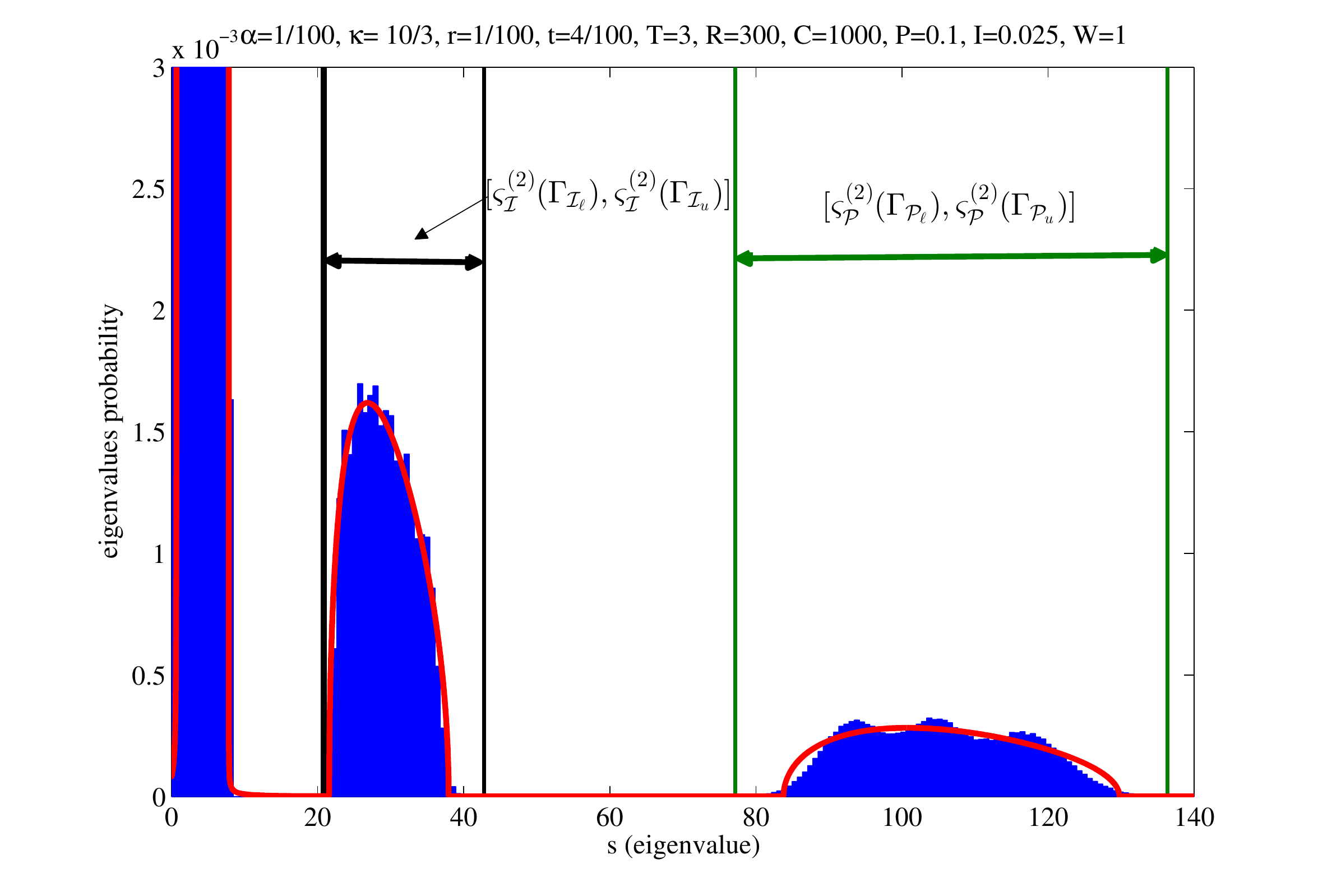,width=\columnwidth}
\caption{\label{fig:boundary_noisy}
Eigenvalue pdf of a network with $T=3,$ $L=2,$ $R=300,$ and $W=0$ with approximation of its support boundaries}
\end{figure}

The tightness of the proposed approximation is assessed by numerical simulations.
In Figure \ref{fig:boundary_noisy} we consider the same communication system as in Figure \ref{fig_asymptotic_evd} but additionally impaired by Gaussian noise with variance equal to 0dB. 
Besides the  histogram of the eigenvalues for a finite system and the asymptotic eigenvalue pdf  drawn in solid line, we show the intervals  $\left[ \varsigma_{\mathcal{I}}^{(2)}(\Gamma_{\mathcal{I}_{\ell}}), \, \varsigma_{\mathcal{I}}^{(2)}(\Gamma_{\mathcal{I}_{u}}) \right]$  and $\left[ \varsigma^{(2)}_{\mathcal{P}}(\Gamma_{\mathcal{P}_{\ell}}), \, \varsigma_{\mathcal{P}}^{(2)}(\Gamma_{\mathcal{P}_{u}}) \right]$. The vertical lines indicate the approximation of the boundaries of the asymptotic pdf obtained by perturbation analysis.  
The approximations based on the second order Taylor expansion $\varsigma^{(2)}_{\mathcal{P}}(x)$ and $\varsigma^{(2)}_{\mathcal{I}}(x)$  include the actual asymptotic support. 

\section{Numerical Results}
\label{numerics}

\begin{figure}
\centerline{\epsfig{file=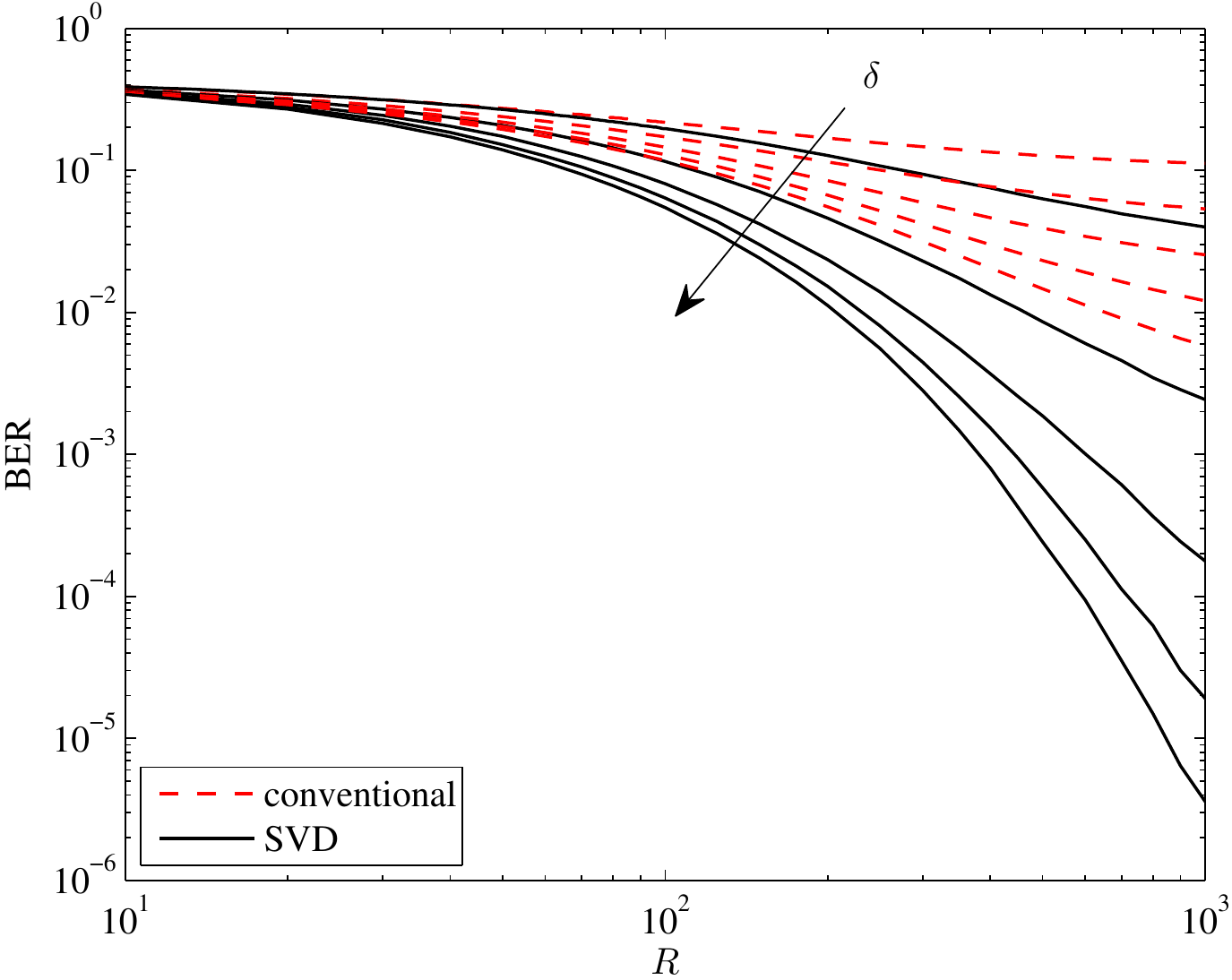,width=.7\columnwidth}}
\caption{\label{BERvsR}
BER vs.\ number of receive antennas with 
$T=5$, $C=100$, $L=6$, $P/W=0.1$ (SNR is $-10$~dB), and the interference distribution $I_k=\frac{P(k{\rm mod} T)}{\delta T}, \forall k=1\dots T$ for various values of the parameter $\delta$ incrementing from 2 to 6 in direction of arrow.}
\end{figure}

In this section, we provide simulation results for the uncoded bit error rate (BER) and compare the proposed SVD-based algorithm, with the conventional linear channel and data estimation scheme considered in \cite{marzetta:10}.  For all cases we set $P/W = 0.1$ (SNR is $-10$~dB), that is, assume that the system operates in the low SNR region. An identical set of orthogonal pilot sequences of length $T$ is adopted by all
the access points to facilitate channel estimation.
We consider first the effect of increasing the number of receive antennas while the rest of the parameters are fixed to $T=5,$ $L=6,$ and $C=100$. As may be observed from Fig.~\ref{BERvsR}, the proposed algorithm (SVD) widely outperforms the receiver based on linear channel estimation in \cite{marzetta:10} (conventional).  Furthermore, it is evident that the proposed algorithm benefits from increased number of receive antennas, irrespective of whether the number of receive antennas is greater or smaller than the coherence time measured in symbol intervals. 

\begin{figure}
\centerline{\epsfig{file=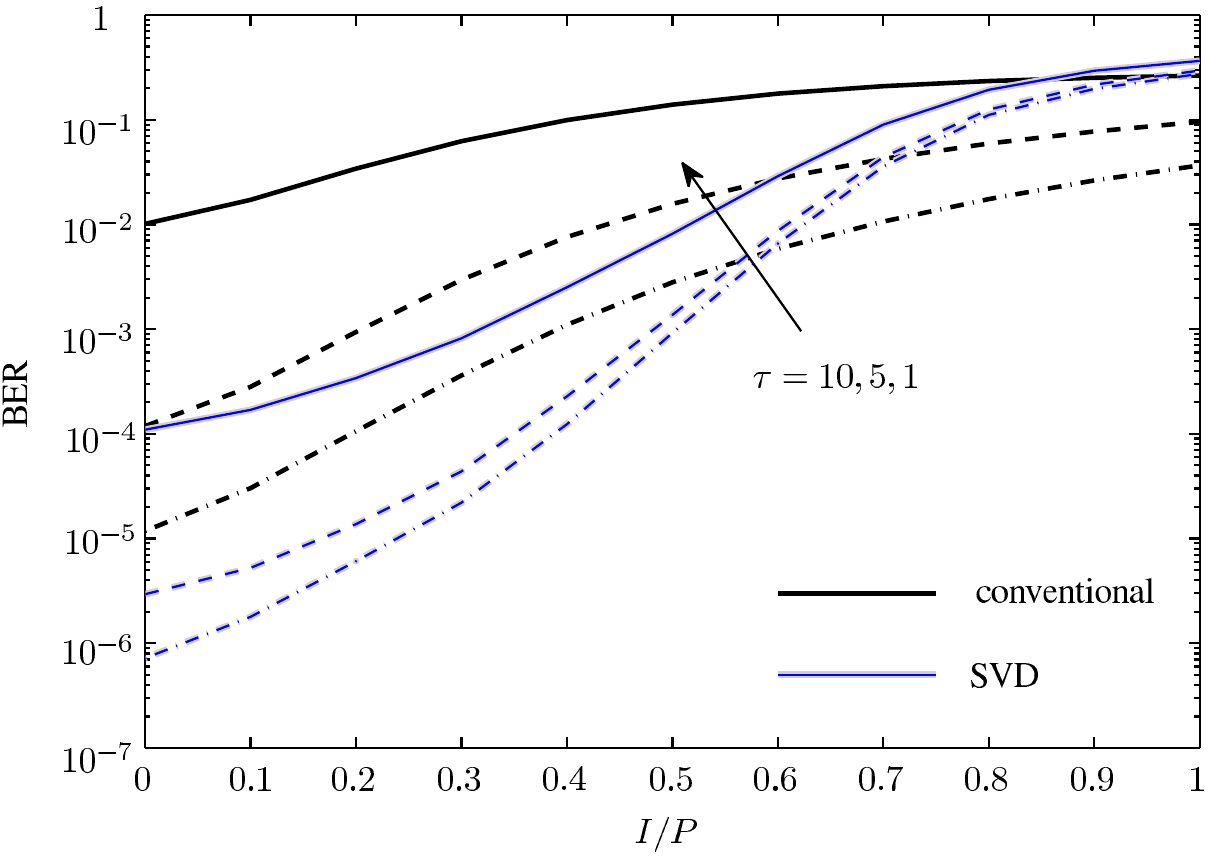,width=.7\columnwidth}}
\caption{\label{BERvsIP}
BER vs.\ relative interference strength with
$T=3,$ $R=300,$ $C=1000,$ $L=2,$ and $P/W=0.1$ (SNR is $-10$~dB).  The number of length-$T$ pilot blocks is $\tau = 1,5,10$.}
\end{figure}

The effect of relative interference strength $I/P$ and number of length-$T$ pilot sequences $\tau$ is plotted in Fig.~\ref{BERvsIP}.  
For $\tau=1$ the same orthogonal pilots are used for all access points.  In the case $\tau = 5,10$, random pilot sequences and zero-forcing channel estimation is employed.  
The RMT thresholds for the given parameters are $I/P=0.61$ and $I/P=0.78$ according to \eqref{RMTthreshold} and \eqref{eq:separability_region_boundary}, respectively.
The proposed algorithm achieves significant performance gains below the RMT thresholds when compared to linear channel estimation.   For very strong interference, however, the conventional receiver outperforms the subspace approach.  The reason is because we always select only the $T$ strongest eigenvectors for projection, but for finite system sizes and close to the RMT threshold this is suboptimal and we lose a large amount of useful signal while projecting towards interference.  This effect can be mitigated by selecting more than $T$ eigenvectors for subspace projection when $I/P$ is expected to be close to the threshold predicted by RMT.

\section{Summary and Conclusions}
\label{conclusions}
We proposed a practical algorithm with polynomial complexity to mitigate pilot contamination in cellular systems with power controlled handoff.
The dominant complexity of this algorithm is a singular value decomposition of the received signal block.
The algorithm was analyzed by means of random matrix theory. 
It was found that the algorithm works well, if certain constraints relating the number of antennas to the coherence time and the signal-to-interference ratio are fulfilled.
Simulations show that under that conditions, the algorithm significantly outperforms linear channel estimation.

This paper has focussed solely on the reverse link channel. 
For the forward link channel, one can exploit channel reciprocity in time-division duplex systems.
Similar to the reverse link channel, knowledge of the full channel matrix is not required. 
Basic considerations of linear algebra show that it is sufficient to know the subspace which the channel vectors of interest span in order to solely require accurate channel estimates for the projected channel \eqref{projectedchannel}.

\begin{appendices}

\section{Eigenvalue Distribution}\label{app:stieltjes_transform}

Consider the random matrix
\begin{equation}
\label{freeconv}
\matr D = \sum\limits_{k=-1}^K a_k \matr B_k \matr C_k
\end{equation}
with $a_k \in \RR$, $\matr B_k\in \CC^{n\times m_k}$ and $\matr C_k\in \CC^{m_k\times n}$ being  random matrices with iid.\ zero-mean entries with variance $1/m_k$ and $1/n$, respectively.
From \cite[Eq.~(35)]{mueller:13a}, we have
\begin{align}
\frac1{{\rm G}_{\matr {DD}^\dagger} (s)} &={-s- \sum\limits_{k=-1}^K \frac {a_k^2 \rho_k  s\,{\rm G}_{\matr {DD}^\dagger}(s)}{\rho_k-a_k^2 s\,{\rm G}_{\matr {DD}^\dagger}^2(s)}}
\label{ll1}
\end{align}
with ${\rm G}_{\matr {DD}^\dagger}(s)$ denoting the Stieltjes transform of the asymptotic eigenvalue distribution of $\matr {DD}^\dagger$.

Next we will distinguish two cases: $\beta\ge 1$ and $\beta\le 1$ .
For $\beta\le 1$, define the matrix $\matr E \in \CC^{\beta n \times n}$ by the decomposition
\begin{equation}
\matr D = \left[
\begin{array}c
\matr E\\
\matr F
\end{array}\right].
\end{equation}
From \cite[Theorem 14.10]{nica:06}, we have
\begin{equation}
\label{origR}
{\rm R}_{\matr {EE}^\dagger}(w) = {\rm R}_{\matr {DD}^\dagger}(\beta w).
\end{equation}
For $\beta\ge 1$, define the matrix $\matr E \in \CC^{\beta n \times n}$ as
\begin{equation}
\matr E = \left[
\begin{array}c
\matr D\\
\matr G_0 \matr P\\
\matr G_1\\
\matr \vdots\\
\matr G_{\lfloor \beta -1\rfloor}
\end{array}\right].
\end{equation}
with the family $(\{\matr D\}, \{\matr G_0\}, \dots, \{\matr G_{\lfloor \beta -1\rfloor}\}, \{\matr P\}) $ being asymptotically free, $\matr G_i \in \CC^{n\times n}$ and ${\rm R}_{\matr G_i \matr G_i^\dagger}(w) =  {\rm R}_{\matr {DD}^\dagger}(w)$ for all $i$, and $\matr P \in \{0,1\}^{n\times (\beta - \lfloor \beta \rfloor)}$ be diagonal with full rank.
 From \cite[Theorem 14.10]{nica:06}, we have
\begin{equation}
\label{recip}
{\rm R}_{\matr {DD}^\dagger}(w) = {\rm R}_{\matr {EE}^\dagger}(w/\beta).
\end{equation}
Note that \eqref{recip} is equivalent to \eqref{origR}.
Thus, we can unify the two cases and continue with \eqref{origR} for any $\beta$.

In the Stieltjes domain, \eqref{origR} translates into
\begin{equation}
\beta {\rm G}_{\matr {EE}^\dagger}(s) = {\rm G}_{\matr {DD}^\dagger}\left(s+\frac{\beta-1}{\beta {\rm G}_{\matr {EE}^\dagger}(s)}\right).
\end{equation}
Together with \eqref{ll1}, we find  an expression similar to \cite[Eq.~(39)]{mueller:13a} which simplifies to 
\begin{align}
s{\rm G}_{\matr {EE}^\dagger\!} \left(s\right)
&=-1-\!\!
 \sum\limits_{k=-1}^K \!\frac {a_k^2 \rho_k \left(s+\frac{\beta-1}{\beta {\rm G}_{\matr {EE}^\dagger\!}(s)}\right)\beta{\rm G}_{\matr {EE}^\dagger\!}^2\left(s\right)}{\rho_k-a_k^2
 \left(s\!+\!\frac{\beta-1}{\beta {\rm G}_{\matr {EE}^\dagger\!}(s)}\right)
 \beta^2{\rm G}_{\matr {EE}^\dagger\!}^2\left(s\right)}.
\end{align}
Now, we consider the matrix $\matr Y$ in \eqref{newmodel} as a special case of $\matr E$.
This implies
\begin{align}
K&=LT\\
\beta & = \frac RC = \frac1\kappa\\
\rho_{-1} & = \frac TC = \frac\alpha\kappa\\
a_{-1}^2 & = PTC \\
\rho_0 & \to \infty\\
a_0^2 &= WC\\
\rho_k & = \frac 1C  \qquad \forall k>0
\\
a_k^2 &= I_{k} C \qquad \forall k>0
\end{align}
and \eqref{fixedpoint} is obtained in the limit $K\to\infty$.
Note that the entries of $\matr B_0\matr C_0$ become iid.\ as $\rho_0\to\infty$.

\section{Eigenvalue Repulsion}\label{app:eigenvalue_repulsion}
\label{app:shift}

In order to find the support of the asymptotic eigenvalue distributions we follow \cite[Eq. (48)]{zee:96}.
There it is shown that the boundaries of the support of the asymptotic eigenvalue distribution are extrema of the inverse of the Stieltjes transform. For a particular example different from ours, the procedure is explained in greater detail in \cite[Chapter 7]{debbah:11}.

Consider the random matrix
\begin{equation}\label{eq:noiseless_system_mat}
\matr E= \left( \sqrt P \matr { A B} + \sqrt I \matr {CD}\right)  \left(  \sqrt P \matr {A B} + \sqrt I\matr {CD}\right) ^\dagger
\end{equation}
where $\matr A\in\CC^{R\times \alpha R}$, $\matr B\in\CC^{\alpha R \times \kappa R}$,  $\matr C\in\CC^{R\times \beta R}$, and $\matr D\in\CC^{\beta R \times \kappa R}$ with iid.\ zero-mean unit-variance entries.
We have from \eqref{fixedpoint} that the limiting Stieltjes transform of $\matr {EE}^\dagger$ obeys
\begin{equation}\label{eq:noiseless_system_stieltjes}
s_\alpha G+1+\frac{\alpha(s_\alpha G+1-\kappa)G}{r \kappa - (s_\alpha G+1-\kappa)G}+\frac{\beta(s_\alpha G+1-\kappa)G}{t \kappa - (s_\alpha G+1-\kappa)G}=0
\end{equation}
with 
\begin{align}
r&= \frac \alpha{PTC}=\frac1{PRC}\\
t&= \frac \alpha{ITC}=\frac1{IRC}.
\end{align}

Solving for $s_\alpha$ leads to a cubic equation and is a tedious task. Nevertheless, Maple 16 can do it symbolically.
For $\alpha=0$, we get
\begin{align}
s_0 &= \frac{ G\kappa -2G +G\beta +t\kappa }{2 G^2} 
-\frac{\sqrt{\beta^2 G^2 + 2\beta Gt\kappa - 2\beta G^2 \kappa + \kappa^2 (G+t)^2} }{2 G^2}.
\label{sexplicitebis}
\end{align}
At the interval boundaries,
\begin{equation}
\frac{\partial s_\alpha}{\partial G} = \frac{Z(G)}{N(G)}
\end{equation}
with obvious definition of the enumerator $Z(G)$ and the denominator $N(G)$, must vanish.
For $\alpha\to 0$, one of the bulks will disappear. Thus, $N(G)$ and $Z(G)$ will have a common zero in the limit $\alpha\to 0$.
This common zero corresponds to the position of the vanishing bulk. Instead of inspecting the zeros of $Z(G)$ when searching for the interval boundary of the vanishing bulk, we can also look at the zeros of $N(G)$\footnote{This procedure is necessary since $Z(G)$ fills many pages even in the limit of $\alpha\to 0$ and finding its zeros is intractable.  However, $\lim_{\alpha\to0} N(G)$ only fills several lines and Maple can find its zeros in closed form.}.
We find that
\begin{equation}
\lim\limits_{\alpha\to0} N(G)=0
\end{equation}
has the following four solutions
\begin{align}
G_1 & = \frac{-\kappa t }{(\sqrt\kappa+\sqrt\beta)^2}\\
G_2 & = \frac{-\kappa t }{(\sqrt\kappa-\sqrt\beta)^2}\\
G_3 & = 0\\
G_4 & = \frac {r\kappa(t-r)}{\kappa (r-t)-\beta r}.
\end{align}
Obviously, $G_4$ is the desired zero, since the other zeros do not depend on $r$.
Plugging into \eqref{sexplicitebis} gives
\begin{align}
s_0(G_4) & = \frac{(t-r+\frac{\beta r}{\kappa})(t-r+\beta r)}{r (t-r)^2 }.
\end{align}
Thus, the presence of interference scales the signal of interest by a factor of 
\begin{equation}
\label{s0G4new}
\frac{s_0(G_4)}{s_0(G_4)\big|_{\beta=0}}=\left(1+\frac{{\beta }/{\kappa}}{t/r-1}\right)\left(1+\frac{{\beta }}{t/r-1}\right).
\end{equation}
The scale factor of the interference is obtained by exchanging the role of signal and interference.

In order to obtain the scale factor for the white noise, we note that for infinite load the interference becomes white.
Thus, we take the limit $t,\beta\to\infty$ with $\zeta=\frac\beta t$ in \eqref{s0G4new} and obtain
\begin{align}
s_0^\infty(G_4) & = \left(\zeta+\frac1r\right) \left(1+\frac {r\zeta }\kappa \right).
\label{s0G4}
\end{align}
Without noise, i.e.\ $\zeta=0$, the signal of interest would be positioned at $1/r$.
Thus, the presence of noise scales the signals by a factor of 
\begin{equation}
\frac{s_0^\infty(G_4)}{s_0^\infty(G_4)\big|_{\zeta=0}}=\left(1+r\zeta\right) \left(1+\frac{r\zeta }\kappa\right).
\end{equation}


\section{The noiseless system}\label{app:noiseless_system_finite_alpha}
\begin{figure}
\centerline{\epsfig{file=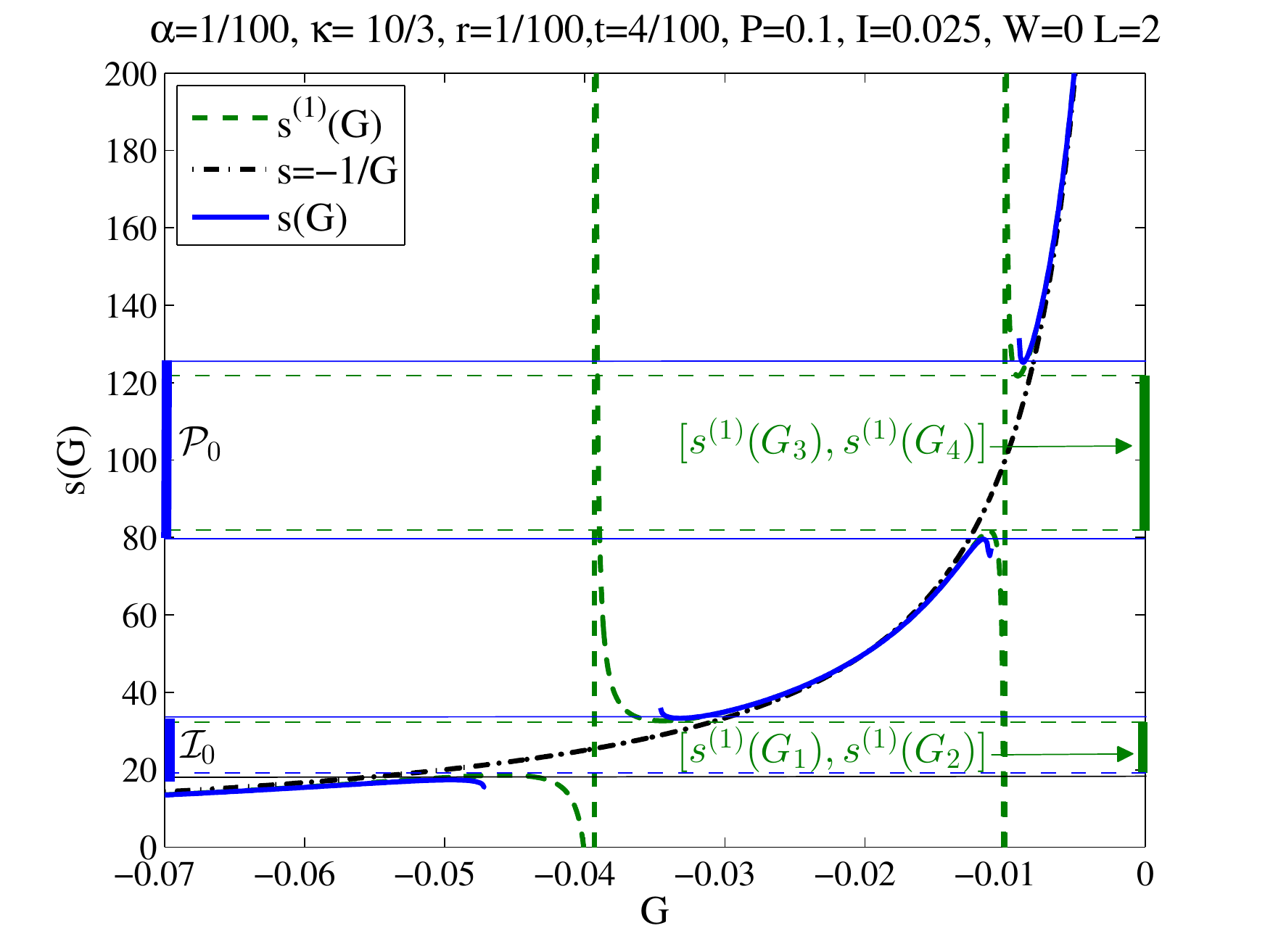,width=.8\columnwidth}}
\caption{\label{fig:various_inverse_approx_noiseless}
Analysis of the inverse Stieltjes transform $s(G),$ its version without perturbation, i.e. $s=-\frac{1}{G},$ and its approximation via perturbation theory $s^{(1)}(G).$ Eigenvalue pdf support of a noiseless system with $L=2, \alpha=\frac{1}{100},$ $\kappa=\frac{10}{4}, $ $P=0.1$ and $I=\frac{P}{4}$ and its approximation by the estimation $[s^{(1)}(G_1), s^{(1)}(G_2)] \bigcup [s^{(1)}(G_3), s^{(1)}(G_4)].$}
\end{figure}

In this section we analyze the behaviour of the noiseless system when the number of interfering signals and signals of interest are proportional and very small compared to the number of receive antennas but not vanishing, i.e. $\alpha \rightarrow 0$. We still consider  the random matrix in (\ref{eq:noiseless_system_mat}) but both the dimensions of the interference and signal subspace grow proportionally, i.e., $\beta=\alpha L.$ Under these assumptions, (\ref{eq:noiseless_system_stieltjes}) can be written as
\begin{align}\label{eq:noiseless_system_stieltjes_perturbed}
\frac{\alpha G(s G+1-\kappa)( \kappa (t+Lr) - (L+1)(s G+1-\kappa)G)}{(r \kappa - (s G+1-\kappa)G) (t \kappa - (s G+1-\kappa)G)}+&\nonumber\\ \frac{(s G+1)(t \kappa - (s G+1-\kappa)G)(r \kappa - (s G+1-\kappa)G)}{(r \kappa - (s G+1-\kappa)G) (t \kappa - (s G+1-\kappa)G)}&=\,0.
\end{align}
By simple inspection, we observe that the numerator $N(s)$ of the l.h.s. in (\ref{eq:noiseless_system_stieltjes_perturbed}) is a function obtained by perturbation of a cubic function in $s$
\begin{equation}
N_0(s)=(s G+1)(t \kappa - (s G+1-\kappa)G)(r \kappa - (s G+1-\kappa)G)
\end{equation}
by a quadratic function in $s$ proportional to $\alpha$
\begin{equation}
N_p(s)=G(s G+1-\kappa)( \kappa (t+Lr) - (L+1)(s G+1-\kappa)G).
\end{equation}
Then, for small $\alpha$ the zeros of the original numerator $N(s)=N_0(s)+\alpha N_p(0)$ can be computed as a perturbed version of the zeros in $N_0(s)$ given by
 \begin{align}\label{eq:zeros_N0}
    s_{0,0}&= -\frac{1}{G}\\
    s_{0,\mathcal{P}}&= -\frac{(1-\kappa)G-\kappa r}{G^2} \\
    s_{0,\mathcal{I}}&=-\frac{(1-\kappa)G-\kappa t}{G^2}.
 \end{align}
Let us observe that (\ref{eq:zeros_N0}) corresponds to the Stieltjes transform of a pdf $p(x)=\delta(x),$ i.e. the eigenvalue distribution of a matrix with all zero eigenvalues and we are interested in its perturbed version by the signal and interference subspaces. Then, we focus on the perturbation of this function to determine the inverse Stieltjes transform. This initial observation will avoid further discussions on the selection of the multiple zeros of $N(s).$ Then, a first order Taylor expansion of $N(s)$ in $s_{0,0}$
\begin{equation}
    N(s) \approx N_p(s_{0,0}) + \left. \frac{\partial N(s)}{\partial s} \right|_{s=s_{0,0}} (s-s_{0,0})
\end{equation}
yields a linear equation in $s$ to determine the approximation of the inverse Stieltjes transform $s^{(1)}(G)$
\begin{equation*}
    s^{(1)}(G)=s_{0,0}+\frac{N_p(s_{0,0})}{\left. \frac{\partial N(s)}{\partial s} \right|_{s=s_{0,0}}}
\end{equation*}
presented in (\ref{eq:inverse_stieltjes_1_order_appr}). Note that $s^{(1)}(G)$ maintains the pole in $G=0$ as the Stieltjes transform of $p(x)=\delta(x)$ but also presents two additional poles in (\ref{eq:poles_s_alpha_1})
as  effect of the perturbation. In Figure \ref{fig:various_inverse_approx_noiseless} we show the exact inverse Stieltjes transform in solid blue lines and compare it with $s^{(1)}(x),$ the approximation via perturbation theory, and $s=-\frac{1}{G}.$ In Figure (\ref{fig:various_inverse_approx_noiseless}), the gaps of the solid blue lines correspond to regions where  $s(G)$ assumes complex conjugate values for real values of $G.$ The extremes of the function $s(G)$ determine the support  $\mathcal{P}_0 \cup \mathcal{I}_0$ of the asymptotic eigenvalue distribution of $\matr{YY}^{\dagger}$ while the extremes of $s^{(1)}(G)$ are related to the estimation $[s^{(1)}(G_1),s^{(1)}(G_2)]\cup[s^{(1)}(G_3),s^{(1)}(G_4)].$ The presence of poles in $s^{(1)}(G)$ is an artefact of the first order Taylor expansion of the polynomial $N(s)$ and corresponds to the region where the  $N(s)$ has two complex conjugate solutions.

\begin{figure}
\centerline{\epsfig{file=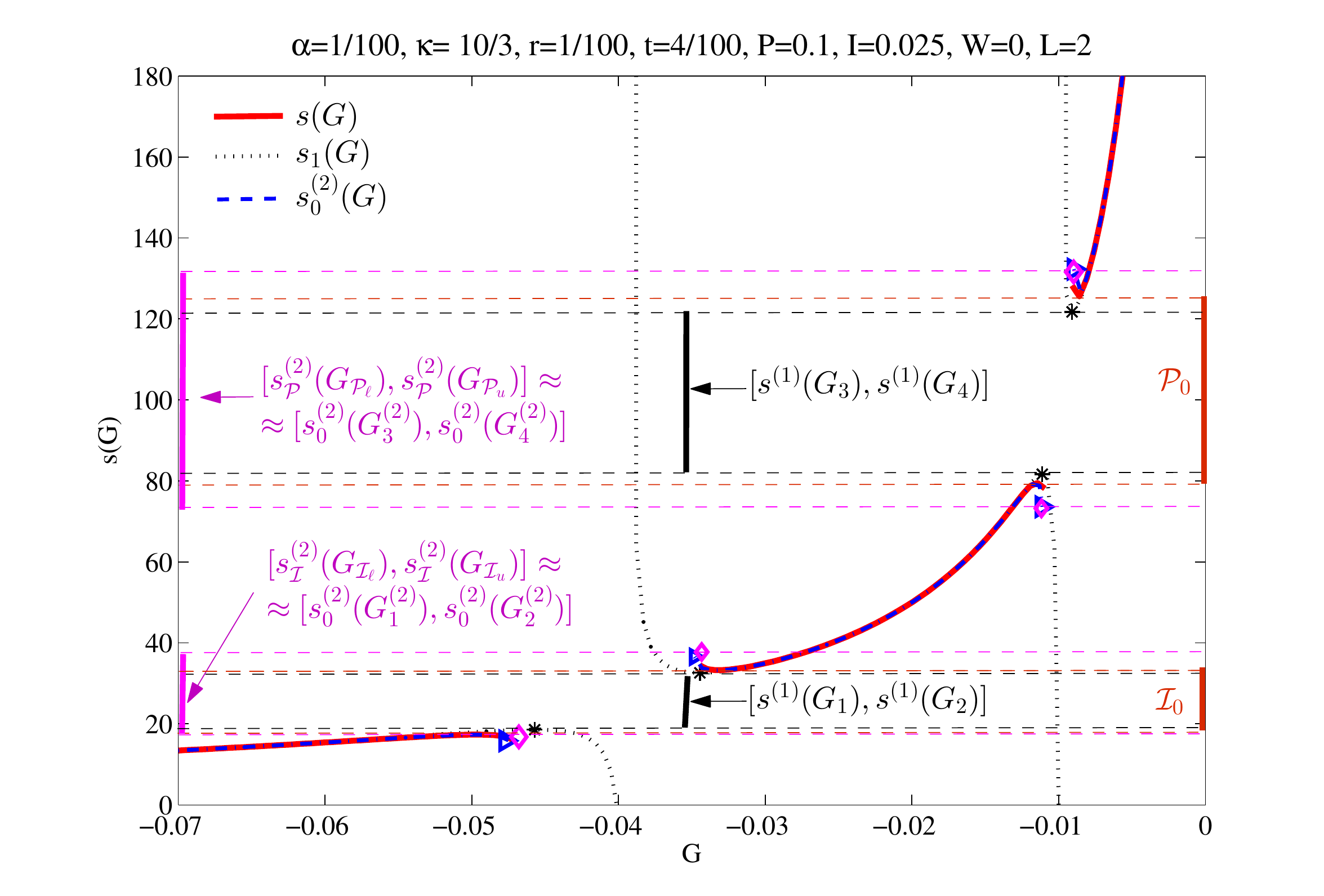,width=.9\columnwidth}}
\caption{\label{fig:comparison_approximation_support_general}
Comparison of the inverse Stieltjes transform $s(G)$ with its approximations $s^{(1)}(G)$ and $s^{(2)}_{0}(G)$ for the case of a noiseless system with $L=2, \alpha=\frac{1}{100},$ $\kappa=\frac{10}{4}, $ $P=0.1$ and $I=\frac{P}{4}.$   The star markers show the extremes of the function $s^{(1)}(G).$ The triangle markers show the points where $s_{0}^{(2)}(G)$ becomes complex for real values of $G.$  The diamond markers show the points where $s_{\mathcal{P}}^{(2)}(G)$ and $s_{\mathcal{I}}^{(2)}(G)$ become complex for real values of $G,$ i.e. in ascending order $G_{\mathcal{I}_{\ell}},$ $G_{\mathcal{I}_{u}},$ $G_{\mathcal{P}_{\ell}},$ and $G_{\mathcal{P}_{u}}.$}
\end{figure}

In order to improve the approximation of the zeros of $N(s)$ in the intervals where they are complex, we consider a second order Taylor expansion of $N(s)$ around $s_{0,0}$ and we obtain the quadratic function in $s:$
\begin{align}
N_0^{(2)}(s)=\,&\big((-2\kappa-\alpha(L+1))G^4-\kappa(t+r)G^3\big)s^2 +\big((\kappa^2+\kappa(2(1+L)\alpha\!-\!4)-2(1+L)\alpha)G^3\nonumber \\
&+\!(\!(r\!+\!t)\kappa^2 +\!(\!(t\!+\!Lr)\alpha\!-\!2t\!-\!2r)\kappa)G^2\!+\!\kappa^2rtG\big)s +(\kappa^2(1-\alpha(L+1))+2\kappa(\alpha(L+1)-1)\nonumber \\
&-\alpha(L+1))G^2+(\kappa^2(r+t-(Lr+t)\alpha) +(\alpha(t+Lr)-t-r)\kappa)G+\kappa^2 rt \label{eq:second_order_approx_s01}
\end{align}
which is a polynomial in $s$ with two zeros
\begin{equation*}
    \widetilde{s}_0^{(2)}(G)=\phi_{0}\pm\rho_0(G)
\end{equation*}
where $\rho_0(G)$ and $\phi_0(G)$ are defined in (\ref{def:rho_0}) and (\ref{def:phi_0}), respectively. The inverse of the Stieltjes transform, selected as perturbation of $s=-\frac{1}{G},$ is $s_{0}^{(2)}(G)$ as defined in (\ref{eq:inverse_stieltjes_2_order_appr}).
Note that $\phi_{0}(G)-\rho_{0}(G)$ cannot be the desired inverse in the interval $[G_{-}^{\infty}, \, G_{+}^{\infty}]$ since it presents a pole in $G=-\frac{(t+r)\kappa}{2 \kappa+\alpha(L+1)}\in [G_{-}^{\infty}, \, G_{+}^{\infty}]$ while $\phi_{0}(G)+\rho_{0}(G)$ does not. However, $\phi_{0}(G)+\rho_{0}(G)$ is not the desired inverse outside the interval $[G_{-}^{\infty}, \, G_{+}^{\infty}]$ since it behaves like $G^{-2}$ in a surrounding of $G=0$ and like $G^{-1}$ for $G \rightarrow \pm \infty.$

In contrast to the analogous problem with the first order Taylor approximation $s^{(1)}(G),$ the computation of the extremes of the function $s_0^{(2)}(G)$ do not have a closed form solution since their computation implies the solution of two polynomial equations of degree seven. In order to acquire deeper insight on the problem, let us observe the behaviour of $s_0^{(2)}(G)$ shown in Figure \ref{fig:comparison_approximation_support_general}. The match between $s(G)$ and $s_0^{(2)}(G)$ is nearly perfect in the surroundings of the extremes. Figure \ref{fig:comparison_approximation_support_general} suggests to approximate the extremes of $s_0^{(2)}(G)$ by the points where $s_0^{(2)}(G)$ becomes complex, i.e. the zeros of $\rho_0(G).$ This approximation implies again the solution of a polynomial equation of degree four yielding the zeros $G_1^{(2)},$ $G_2^{(2)},$ $G_3^{(2)},$ and $G_4^{(2)}$ with $G_1^{(2)}\leq G_2^{(2)} \leq G_3^{(2)} \leq G_4^{(2)}.$ Although the zeros $G_1^{(2)},$ $G_2^{(2)},$ $G_3^{(2)},$ and $G_4^{(2)}$ can be expressed in closed form, their expression is too cumbersome to be insightful. In order to obtain more practical and useful results we consider the second order Taylor expansion of $N(s)$ in $s_{0,\mathcal{P}}$ and $s_{0,\mathcal{I}},$ which yields
\begin{align}
N(s)\approx \,&\left((\kappa-\alpha(L+1))G^4+\kappa(2r-t)G^3\right)s^2+\big((2\kappa(\alpha(L+1) +1-\kappa)-2\alpha(L+1))G^3\nonumber \\ &+((t-5r)\kappa^2+\kappa (4r-2t+\alpha t+\alpha L r))G^2+\kappa^2r(t-3r)G\big)s +(\kappa^2(-2-L\alpha-\alpha+\kappa)\nonumber \\
&-\alpha(L+1)+\kappa(2\alpha(L+1)+\kappa))G^2  +(\kappa^2(3\kappa r-\alpha (L r+t)+t-5r)+\kappa(2-t)\nonumber \\
&+\alpha \kappa(t+Lr))G+3\kappa^2r(\kappa r-r+\kappa^2 t+\frac{\kappa r^2}{G}) \label{eq:s0P_second_order_approx} \\
N(s) \approx\,& \left((\kappa-\alpha(L+1))G^4+\kappa(2t-r)G^3\right)s^2+\big((-2\kappa^2+2(1+(1+L)\alpha)\kappa-2\alpha(L+1))G^3\nonumber \\
 &+((r-5t)\kappa^2+((t+Lr)\alpha-2r+4t)\kappa)G^2+\kappa^2(r-3t)t G\big)s +(\kappa^3-(2+(L+1)\alpha)\kappa^2\nonumber \\
 &+(1+2(1+L)\alpha)\kappa-(L+1)\alpha)G^2+(3\kappa^3 t+(r-(Lr+t)\alpha-5t)\kappa^2  +((t+Lr)\alpha\nonumber \\ 
 &-r+2t)\kappa)G+3\kappa^3t^2+t(r-3t)\kappa^2+\frac{t^3\kappa^3}{G} 
 \label{eq:s0I_second_order_approx}.
\end{align}
The zeros of (\ref{eq:s0P_second_order_approx}) and (\ref{eq:s0I_second_order_approx}) are relatively good approximations of the actual inverse Stieltjes transform $s(G)$  in the surrounding of the poles (\ref{eq:poles_s_alpha_1}). Let us denote them $\widehat{s}^{(2)}_{\mathcal{P}}(G)$ and  $\widehat{s}^{(2)}_{\mathcal{I}}(G),$ respectively. By using again as approximation for the extremes of $\widehat{s}^{(2)}_{\mathcal{P}}(G)$ and  $\widehat{s}^{(2)}_{\mathcal{I}}(G),$  the values of $G$ where $\widehat{s}^{(2)}_{\mathcal{P}}(G)$ and  $\widehat{s}^{(2)}_{\mathcal{I}}(G)$ become complex, i.e. the points where the discriminants of (\ref{eq:s0P_second_order_approx}) and (\ref{eq:s0I_second_order_approx}) vanish, we can obtain simpler approximations of the extremes.  In fact, the discriminants of (\ref{eq:s0P_second_order_approx}) and (\ref{eq:s0I_second_order_approx}) are again quartic polynomials in $G$ but with two zeros in $G=0.$ The other two zeros can be easily computed and are given by (\ref{eq:approx_estremes_P})
for  $\widehat{s}_{\mathcal{P}}^{(2)}(G)$  and by (\ref{eq:approx_estremes_I}) for $\widehat{s}_{\mathcal{I}}^{(2)}(G).$ Then, observing that the irrational components of $\widehat{s}_{\mathcal{P}}^{(2)}(G)$ and $\widehat{s}_{\mathcal{I}}^{(2)}(G)$ vanish in $G=G_{\mathcal{P}}^{(2)}$ and $G=G_{\mathcal{I}}^{(2)},$ respectively, $\widehat{s}_{\mathcal{P}}^{(2)}(G_{\mathcal{P}}^{(2)})={s}_{\mathcal{P}}^{(2)}(G_{\mathcal{P}}^{(2)})$ and $\widehat{s}_{\mathcal{I}}^{(2)}(G_{\mathcal{I}}^{(2)})={s}_{\mathcal{I}}^{(2)}(G_{\mathcal{I}}^{(2)})$ with ${s}_{\mathcal{P}}^{(2)}(G)$ and ${s}_{\mathcal{I}}^{(2)}(G)$ defined in (\ref{eq:s_P^(2)}) and (\ref{eq:s_I^(2)}), respectively. The observation that the instances of $G_{\mathcal{P}}$ and $G_{\mathcal{I}}$ with sign plus are not greater than the corresponding instances with sign minus, i.e. $G^{(2)}_{\mathcal{P}_{\ell}}\leq G^{(2)}_{\mathcal{P}_{u}}$ and $G^{(2)}_{\mathcal{I}_{\ell}}\leq G^{(2)}_{\mathcal{I}_{u}},$  yields the approximations (\ref{eq:approx_P0}) and (\ref{eq:approx_I0}).

By appealing to the previous results, in the following we derive condition (\ref{eq:separability_region_boundary}) for bulk separability.

Under the assumptions of physical interest that $L\in \mathbb{N}^{+}$ and $t \geq r\geq 0,$ $G_{\mathcal{P}}^{(2)}$ and $G_{\mathcal{I}}^{(2)}$ are all negative real zeros\footnote{These conditions are obtained by enforcing that the arguments of the square roots in (\ref{eq:s0P_second_order_approx}) and (\ref{eq:s0I_second_order_approx}) are nonnegative.} if
\begin{align}
    \frac{t}{r} &\geq \max \textstyle \left( 1+\frac{\alpha}{2 \kappa}+ \frac{\alpha}{2 \kappa} \sqrt{1+\frac{4L \kappa}{\alpha}}, \frac{1+\frac{L \alpha }{2 \kappa}+ \frac{L \alpha}{2 \kappa} \sqrt{1+\frac{4 \kappa}{L^2 \alpha}}}{1+(L-1) \frac{\alpha}{\kappa}}\right) = 1+\frac{\alpha}{2 \kappa}+ \frac{\alpha}{2 \kappa} \sqrt{1+\frac{4L \kappa}{\alpha}} \label{eq:condition_set_0}
\end{align}
or equivalently
\begin{equation}\label{eq:condition_set_0_bis}
    0 \leq \frac{\alpha}{\kappa} \leq \frac{(t-r)^2}{r(t+(L-1)r)}.
\end{equation}
By simple inspection, it is easy to verify that $G^{(2)}_{\mathcal{I}_{\ell}}\leq G^{(2)}_{\mathcal{P}_{u}}$ under the above mentioned conditions of physical interest. However, it is interesting to determine under which conditions the two intervals $[G^{(2)}_{\mathcal{P}_{\ell}} , G^{(2)}_{\mathcal{P}_{u}}] $ and $[G^{(2)}_{\mathcal{I}_{\ell}} , G^{(2)}_{\mathcal{I}_{u}}] $ do not intersect, i.e. when $G^{(2)}_{\mathcal{I}_{u}}  \leq G^{(2)}_{\mathcal{P}_{\ell}}. $ It can be verified, for example using Maple, that this last condition is satisfied if $\frac{\alpha}{\beta}$ and $\beta=\frac{r}{t}$ satisfy (\ref{eq:separability_region_boundary}). Additionally, condition (\ref{eq:separability_region_boundary}) implies also (\ref{eq:condition_set_0_bis}).

\section{The Noisy System}\label{app:noisy_system_bulk_boundary}
The analysis of the system with noise follows along lines similar to the ones adopted in the previous section. The fixed point equation for the Stieltjes transform of the eigenvalue pdf in (\ref{fixedpoint}) can be rewritten as
\begin{align}\label{eq:noisy_system_stieltjes}
s G+1+\frac{\zeta(sG+1-\kappa) G}{\kappa}
+\frac{\alpha(s G+1-\kappa)G}{r \kappa - (s G+1-\kappa)G} +\frac{\alpha L(s G+1-\kappa)G}{t \kappa - (s G+1-\kappa)G}&=0
\end{align}
with $r,t$ and $\zeta$ defined in (\ref{def:r}), (\ref{def:t}), and (\ref{def:zeta}). 

The inverse function $s(G)$ can be obtained as a zero of the numerator of (\ref{eq:noisy_system_stieltjes}). As in the previous section,  this is a cubic function in $s$ obtained as perturbation of the cubic function
\begin{align}\label{eq:N_0_noisy}
    N_0(s)=\, &\big(G(\kappa+\zeta G)s + \zeta (1-\kappa)G+ \kappa\big)\big(r \kappa - (s G+1-\kappa)G\big)\big(t \kappa - (s G+1-\kappa)G\big)
\end{align}
by a quadratic function 
\begin{equation}\label{eq:N_p_noisy}
    N_{p}(s)=\kappa G(s G+1-\kappa)( \kappa (t+Lr) - (L+1)(s G+1-\kappa)G).
\end{equation}
Simple inspection of (\ref{eq:N_0_noisy}) and (\ref{eq:N_p_noisy}) shows that the introduction of noise has the only effect of  modifying $s_{0,0}$ in (\ref{eq:zeros_N0}) into
\begin{equation}\label{eq:zeros_N_0_noisy}
    \widetilde{s}_{0,0}=-\frac{\zeta(1-\kappa)G+\kappa}{(\kappa+\zeta G) G}
\end{equation}
while, up to a scaling factor $\kappa$, it leaves unchanged the  perturbation $N_p(s).$
The first order Taylor expansion of $N(s)$ in $\widetilde{s}_{0,0}$ yields to an approximation of the inverse Stieltjes transform whose extremes computation requires the solution of a polynomial of degree six and it is not feasible in closed form. Thus, we do not discuss further this case. On the contrary, for the second order expansion, all the results obtained for the noiseless system can be extended. The second order expansion of $N(s)$ in $\widetilde{s}_{0,0}$ yields a polynomial whose discriminant is again a quartic equation in $G.$ Similarly to the  noiseless case, to obtain approximations of practical use we consider the second order expansions of $N(s)$ in $s_{0,\mathcal{P}}$ and $s_{0,\mathcal{I}}$ and we approximate the extremes of the inverse Stieltjes transform by the zeros of the corresponding discriminants. The expansion in $s_{0,\mathcal{P}}$  yields the approximations  $\Gamma_{\mathcal{P}_{\ell}}^{(2)}$  and $\Gamma_{\mathcal{P}_{u}}^{(2)}$ in (\ref{eq:Gamma_P}) for the extremes of the inverse Stieltjes transform and the approximation of the inverse Stieltjes transform boils down to $\varsigma_{\mathcal{P}}^{(2)}(\Gamma_{\mathcal{P}_{u}}^{(2)})$ and $\varsigma_{\mathcal{P}}^{(2)}(\Gamma_{\mathcal{P}_{\ell}}^{(2)}),$ with $\varsigma_{\mathcal{P}}^{(2)}(G)$  defined in (\ref{eq:varsigma_P}), when evaluated in  $\Gamma_{\mathcal{P}_{\ell}}^{(2)}$  and $\Gamma_{\mathcal{P}_{u}}^{(2)}.$ Similar considerations hold for approximation based on the second order expansion in $s_{0,\mathcal{I}}.$

\end{appendices}
\bibliographystyle{IEEEtran}
\bibliography{lit}
\end{document}